\documentclass[usenames,dvipsnames,runningheads]{llncs}

\usepackage{graphicx}
\usepackage[nolist]{acronym}% http://ctan.org/pkg/acronym

\usepackage{subfigure}

\usepackage{multicol}
\usepackage{listings}
\usepackage{numprint}
\usepackage{bytefield}
\usepackage[binary-units]{siunitx}

\usepackage{hyperref}

\usepackage{tikz}
\usetikzlibrary{
  arrows,
  automata,
  backgrounds,
  calc,                     % drawing the background after the foreground
  chains,
  decorations.pathmorphing, % noisy shapes
  decorations.pathreplacing,
  fit,                      % fitting shapes to coordinates
  math,
  matrix,
  mindmap,
  patterns,
  positioning,
  scopes,
  shapes.gates.logic.US,
  shapes.geometric,
  shapes.geometric,
  shapes.symbols,
  shadings,
  shadows,
  spy,
  decorations.pathmorphing, % noisy shapes
  trees,
  tikzmark,
}

\newcommand{\colorbitbox}[3]{%
  \rlap{\bitbox{#2}{\color{#1}\rule{\width}{\height}}}%
  \bitbox{#2}{#3}
}

\newcommand{\etal}[1]{\textit{et al.}}

\title{Java Card\texttrademark{} Virtual Machine Memory Organization: a Design Proposal}

\author{Guillaume Bouffard\orcidID{0000-0002-2046-369X} \and Vincent Giraud \and L\'eo Gaspard}

\authorrunning{G. Bouffard \textit{et al.}}

\institute{
  National Cybersecurity Agency of France (ANSSI),\\
  51, bd de La Tour-Maubourg,\\75700 Paris 07 SP, France\\
  \email{guillaume.bouffard@ssi.gouv.fr}}

\begin{document}

\maketitle

\begin{abstract}%
  The \ac{JCVM} platform is widely deployed on security-oriented components.
  \ac{JCVM} implementations are mainly evaluated under security schemes.
  However, existing implementation are close-source without detail. We believe
  studying how to design \ac{JCVM} will improve them and it can be reused by the
  community to improve Java Card security.

  In 2018, Bouffard \etal{}~\cite{conf/sstic/bouffard18} introduced an \ac{OS}
  which aims at running \ac{JCVM} compatible implementation. This \ac{OS} is
  compatible with several \ac{COTS} components. This is a first step to design a
  secure \ac{JCVM} platform.

  However, some important details are missing to design a secure-oriented Java
  Card platform. In this article, we focus on the \ac{JCVM} memory. This memory
  contains everything required to run \ac{JCVM} and applets. Currently,
  \ac{JCVM} memory is out of the Java Card specification and each \ac{JCVM}
  developer use his own approach. Based on the existing tools and
  documentation, we explain how to extract from the Java Card toolchain every
  data required by applets and \ac{JCVM}. When data to store in memory are
  identified, this article introduces how to organize required data onto \ac{JCVM}
  memory.

\end{abstract}

\acresetall{}

\section{Introduction}\label{sec:introduction}

Java Card~\cite{book/oracle15/jcvm_spec} is a Java subset designed to run on
resource-limited devices such as secure components or IoT devices. It implements
a \ac{VM} to run Java applications in a portable and secure environment. Java
Card platforms aim at running security-oriented applications where a high
security level is required both on physical and software layout. Therefore, most
of \ac{JCVM} implementations are evaluated within the Common Criteria
scheme~\cite{ppjavacard}. However, these implementations are closed source
without details.\newline

The Java Card platform implementations security has been studied against
hardware~\cite{conf/cardis/BarbuTG10,conf/cardis/BarbuDH11,conf/secrypt/BarbuHD12,conf/sstic/Dubreuil16}
and
software~\cite{journals/compsec/BouffardL15,conf/cardis/BouffardIL11,conf/cardis/Faugeron13,conf/cardis/MostowskiP08,conf/dbsec/RazafindralamboBL12,journals/istr/HamadoucheL13,conf/cardis/Lancia12,conf/sarssi/HamadoucheBLDNMR12,journals/jce/IdrissiBLH17,journals/virology/Iguchi-CartignyL10,conf/cardis/LanciaB15}
attacks. For every attack, efficient countermeasures are presented. However,
they are implemen\-ta\-tion-dependent. Currently, there is no public reference
implementation to compare these countermeasures. Therefore, we think a \ac{JCVM}
public implementation design where the community can contribute will improve the
Java Card security.\newline

To propose implementation design to a community, one should target a platform
easily accessible to evaluate and improve the existing solutions. For this
purpose, Bouffard \etal{}~\cite{conf/sstic/bouffard18} have introduced an \ac{OS}
designed to support \ac{JCVM} specificity. Their \ac{OS} exposes security
features based on hardware mechanisms embedded on most \ac{COTS} components.
Therefore, the next step would be to design a secure-oriented \ac{JCVM} which
reuses the features provided by this \ac{OS}.\newline

In this article, we focus on how to organize the \ac{JCVM} memory. The \ac{JCVM}
memory contains everything required to run the \ac{JCVM} and Java Card
applications. Therefore, the \ac{JCVM} memory is composed of the Java Card
standard and vendors' \ac{API}, GlobalPlatform manager and applications code and
other data. How to organize the \ac{JCVM} memory is not analysed by the
literature and should be studied to improve the security \ac{JCVM}. Based on the
existing tools and the corresponding documentations, we propose a way to
initialize and organize the \ac{NVM} memory usable by a \ac{JCVM}
implementation.

\subsection{Java Card Platform}\label{sec:java-card-practice}

Java Card technology is a secure-oriented platform designed to be embedded in
resource-limited devices where a high security level is required. To ensure
this, security mechanisms are embedded both inside and outside the \ac{JCVM}. On
the one hand, outside the \ac{JCVM}, organizational security rules aim at
preventing from loading malicious applications. On the other hand, the \ac{JCVM}
embeds several platform security features. The Java Card security model is
described in section~\ref{sec:jc-secure-model}. However, before presenting it,
an overview of the \ac{JCVM} architecture is introduced in
section~\ref{sec:jc-architecture}.

\subsubsection{Java Card Architecture}\label{sec:jc-architecture}

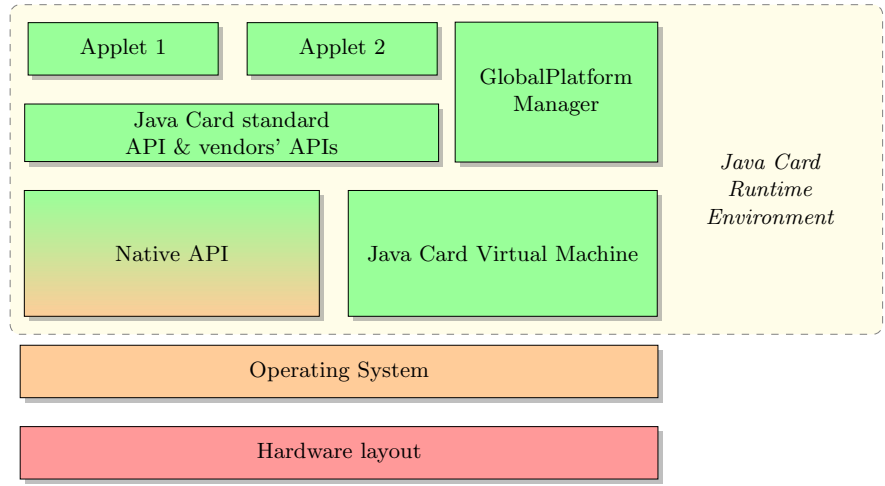
\begin{figure}[!hbt]
  \centering
  \scalebox{0.93}{
    \begin{tikzpicture}[
      start chain=1 going below,
      start chain=2 going right,
      node distance=4mm,
      level/.style={
        on chain=1,
        text centered,
      },
      empty_box/.style={
        rectangle,
        text width=6em,
        align=center
      },
      parts_vm/.style={
        draw,
        rectangle,
        align=center,
        drop shadow,
        fill=green!40,
        text width=4cm,
        text centered,
        minimum height=0.75cm,
      },
      app/.style={
        parts_vm,
        text width=2.5cm
      },
      jcvm/.style={
        parts_vm,
        text width=8.9cm
      },
      jre/.style={parts_vm},
      os/.style={
        jcvm,
        fill=orange!40
      },
      hardware/.style={
        jcvm,
        fill=red!40
      },
      ]

      \node [app, on chain=2]                             (app1)   {Applet 1};
      \node [app, on chain=2, continue chain=going right] (app2)   {Applet 2};

      \node
      [jcvm, xshift=-1.58cm, on chain=2, continue chain=going below, text width=5.7cm] (api)
      {Java Card standard \acs{API} \& vendors' \acsp{API}};

      \node [jcvm, text width=4cm, minimum height=1.8cm, xshift=-0.855cm, shade, top color=green!40, bottom color=orange!40, on chain=2]          (jni)    {Native \acs{API}};

      \node [jcvm, text width=4.2cm, minimum height=1.8cm, on chain=2, continue chain=going right]
      (jcvm) {\acl{JCVM}};

      \node [os, on chain=2, xshift=-2.34cm, continue chain=going below]
      (OS) {\acl{OS}};

      \node [hardware, on chain=2]
      (hard) {Hardware layout};%: CPU + Memories + I/O};

      \path (app2.north east)+(0.25,0) node (app3_1) {};
      \path (api.south -| jcvm.east)+(0,0) node (app3_2) {};
      \draw [draw,
      align=center,
      drop shadow,
      fill=green!40,
      text width=4cm,
      text centered,] (app3_1) rectangle (app3_2)
      node (app3) [pos=.5] {GlobalPlatform\\Manager};

      \path (app3.north east) -- (jcvm.south east) node[midway]
      (jcre) [empty_box, right = of api.east, xshift=-0.5cm, text
      width=2.5cm] {\textit{\acl{JCRE}}};

      \begin{pgfonlayer}{background}
        \path (app1.north west)+(-0.25,0.25) node (a) {};
        \path (jcvm.south -| jcre.east)+(0.25,-0.25) node (b) {};
        \path [fill=yellow!10,rounded corners, draw=black!50, dashed] (a) rectangle (b);
      \end{pgfonlayer}

    \end{tikzpicture}
  }
  \caption{Java Card architecture.}
  \label{fig:jc_architecture}
\end{figure}

The \ac{JCVM} architecture shown in figure~\ref{fig:jc_architecture}, is
inspired by the \ac{JVM}, where each application developed in Java runs onto the
\ac{JCRE}. Each application, named applet, can only execute functions from the
Java Card \ac{API} provided by the platform.

\paragraph{Calling native \ac{API} from Java Card applet}

To call the native code from the Java code, the \ac{JVM} has an interface, named
\ac{JNI}. Unlike \ac{JVM}, calling native codes from Java Card applets is not
allowed by the \ac{JCVM}~\cite[section 3.3]{book/oracle15/jcvm_spec}. There is
no specified path to call this kind of code. However, specific features, like
cryptographic aperations, needs to be run natively for security or efficiency
matters.

Some Java Card platforms such as NXP JCOP allow the native code to be loaded.
This native code is executed in a proprietary sandbox called
SecureBox~\cite{press_release_nxp_2020}. This feature is a proprietary
implementation designed by the platform developers out of any public
specifications.

% Few
% works~\cite{journals/virology/BouffardL14,conf/cardis/LanciaB15,journals/ijisec/MesbahLM19}
% focus on how \ac{JCVM} developers implement calls to native methods process.
% These articles show that the use of undocumented features does not prevent
% privilege escalation. We will reuse information disclosed by these papers to
% improve the calling native methods process.

\paragraph{GlobalPlatform Manager}

At the application level, a content manager is used to strengthen security in
the environment. The GlobalPlatform specification~\cite{book/spec/gp_spe11}
helps system developers who want to implement features widely spread across
the market. Besides content and permission administration, such measures include
platform and applications state management, internal and external communications
protection, and inter-applications services supervision. These features are
often integrated within a daemon, which must always be available at runtime.
GlobalPlatform security managers tend to be useful to platform manufacturers and
operators, since they allow them to specify their own policies and rules
concerning the platform.

\subsubsection{Java Card security model}\label{sec:jc-secure-model}

\begin{figure}[!hbt]
  \centering
  \resizebox{\linewidth}{!}{
  \begin{tikzpicture}[
    frame/.style={
      rectangle,
      draw=black!80,
      fill=blue!20,
      text width = 6em,
      align = center,
    },
    secu/.style={
      frame,
      draw=black!80,
      fill=yellow!20
    },
    conver/.style={
      frame,
      draw=black!80,
      fill=orange!20
    },
    verif/.style={
      frame,
      draw=black!80,
      fill=red!30
    },
    firewall/.style={
      frame,
      text width=4em,
      align=center,
      draw=black!80,
      fill=orange!30
    },
    ]

    \matrix [column sep=0.75cm,row sep=0.75cm] {%
      \node (class) [frame] {Java \texttt{class} files};
      \pgfmatrixnextcell
      \pgfmatrixnextcell
      \node (cap) [frame] {Java Card \acsu{CAP} file};
      \pgfmatrixnextcell
      \node (install) [secu] {Installation module};
      \pgfmatrixnextcell
      % \pgfmatrixnextcell
      \node (app) [firewall] {Installed applet};
      \\
      \node (bcc) [conver] {Bytecode converter};
      \pgfmatrixnextcell
      \node (bcv) [verif] {\acl{BCV} (\acs{BCV})};  \pgfmatrixnextcell
      \node (bcs) [secu]  {Bytecode signature}; \\
      \node (exp) [frame] {Java Card \texttt{export} files};\\
    };

    \path[-latex]
    (class) edge[] (bcc)
    (bcv) edge[] (bcs)
    (bcc) edge[] (bcv)
    (bcs) edge[] (cap)
    (cap) edge[] (install)
    (install) edge[] (app);

    \draw[-latex,dashed] ($(bcc.south)+(0.5em,0)$) -- ($(exp.north)+(0.5em,0)$);
    \draw[latex-] ($(bcc.south)-(0.5em,0)$) -- ($(exp.north)-(0.5em,0)$);

    \node (first dotted box) [dashed, rounded corners, draw, fit = (app)] {};
    \node (label) at (first dotted box.south) [below, inner sep=2mm] {Firewall};

    \draw [draw, dashed, transform canvas={xshift=-1.0em}]
    (cap.north east -| install.north west) -- (exp.south -| install.west);

    \node at (exp.east -| bcv.south) [anchor=west, xshift=-1.0cm, text width=15em] (label_off)
    {\footnotesize \it Java Card security mechanisms out of the \ac{JCVM}.};

    \node at (label_off.east -| install.south) [anchor=west, xshift=-0.9cm, text width=14em]
    {\footnotesize \it Java Card security mechanisms embedded in the \ac{JCVM}.};

  \end{tikzpicture}
  }
  \caption{Java Card security model.}
  \label{fig:jc_security_model}
\end{figure}
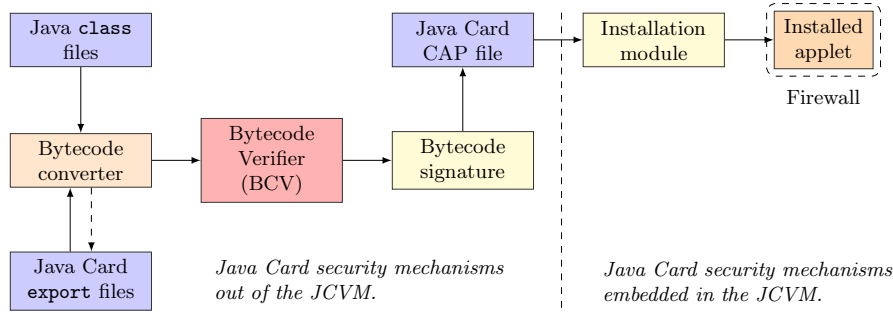

As described in figure~\ref{fig:jc_security_model}, to install an applet on the
Java Card platform, one must first implement it in Java-language and then build
it within Java compiler (\texttt{javac}) to obtain Java \texttt{class} files.
Those \texttt{class} files are not designed to be embedded in a resource-limited
device. Indeed, the Java \texttt{class} files are executed as-is with a link
resolution by name. This kind of resolution is expensive. The adopted solution
was to translate the Java-Class name to token during a step made by the Java Card
converter, included in the public Java Card toolchain\footnote{The Java Card
  toolchain is available on the Oracle's website:
  \url{https://www.oracle.com/java/technologies/java-card-tech.html}}. If
the \texttt{class} file to convert shares features usable by other
applications, a Java Card \texttt{export} file is also generated. The
\texttt{export} file contains, for each Java name element, the associated token
embedded on the device. Therefore, the \texttt{export} files are also used by
the bytecode converter during the translation process. After this translation,
the Java Card files are checked by the \ac{BCV} which statically verifies the
compliance to the Java Card security rules. The translated Java \texttt{class}
files are named the \acsu{CAP} (for \acl{CAP}) file. There is a unique \ac{CAP}
file by converted package, and it is signed to ensure its integrity and
authenticity. On the device, the GlobalPlatform layer verifies the applet
signature. \newline

When the applet \ac{CAP} file is obtained and signed, the applet developer needs
GlobalPlatform loading keys to load their applet or \ac{API}. During the
installation process, an embedded security module checks some security elements.
As this verification is not described by the Java Card
specification~\cite{book/oracle15/jcvm_spec} some works focus on reversing this
process~\cite{journals/virology/BouffardL14,journals/ijisec/MesbahLM19}. After
the applet is installed, it runs in its context segregated by the Java Card
Firewall. It ensures that the applet accesses only its data or specific shared
features.%\newline

% In this section, we showed how the Java Card platform is organized from both
% applet developers and \ac{VM} developers' point of view. Applet developers have
% more specified information about how to make their applet compatible with the
% targeted \ac{JCVM}. However, \ac{JCVM} developers have less specified
% constraints of their \ac{VM} design and, therefore, they are free to
% architecture \ac{VM} internals.

\subsection{Related works}\label{sec:related-works}

Contributions about Java Card mainly focus on about how to protect \ac{JCVM}
implementations by proposing
countermeasures~\cite{conf/snds/RazafindralamboBTL12,journals/ijsse/DubreuilBTL13,prevost2006application,conf/cardis/BouffardIL11,conf/cardis/BarbuDH11,conf/javacard/BieberCMGLWZ00,conf/fgit/SereIL10,conf/aasri/farissiALM13}.
To improve Java Card, few works study how to develop \ac{JCVM} from scratch.

To securely develop a \ac{JCVM}, two approaches have been explored. On the one
hand, some works design hardware blocks regarding Java Card security
requirements. These hardware blocks aim at protecting the runtime environment.
Lackner \etal{} designed mitigation to prevent type confusion onto the
stack~\cite{conf/cardis/LacknerBLWS12,conf/hipeac/LacknerBWS14} and the
heap~\cite{conf/cardis/BouffardLLL14}. In his master's thesis,
Zelle~\cite{masterthesis/Zelle15} describes how to protect the Java Card
firewall through a \ac{MPU}. However, these approaches are based on hardware
mechanisms implemented on FPGA\@. Based on proprietary extensions, these
approaches cannot be used on \ac{COTS} components.

On the other hand, software approaches are also studied to securely design a
\ac{JCVM}. Dufay~\cite{phd/dufray03} has developed a formal proof of the
\ac{JCVM} platform design based in Coq-language able to generate compiled code.
However, his work is based on an old version of Java Card and the developed
formal model is partially available in his thesis and articles; it's not enough
to generate a \ac{JCVM} implementation. Lafer~\cite{masterthesis/Lafer14}
develops a \ac{JCVM} with x86 and 8051 implementations. This work seems
interesting, however, the master thesis suffers from missing implementation
details and no source code is provided. Recently, Bouffard
\etal{}~\cite{conf/sstic/bouffard18} introduce an \ac{OS} in Rust-language to
run a \ac{JCVM} implementation. Their \ac{OS} provides backend to support recent
\ac{JCVM}.

\subsection{Contribution}\label{sec:contribution}

Every listed article in section~\ref{sec:related-works} introduces very little
details on \ac{JCVM} internals. These details are not enough to develop a
\ac{JCVM} and should be developed to improve the Java Card implementation
security. In this article, we address the problem of memory organization usable
by the \ac{JCVM} to store and to manipulate data. This part is a first step to
design a secure-oriented Java Card platform.\newline

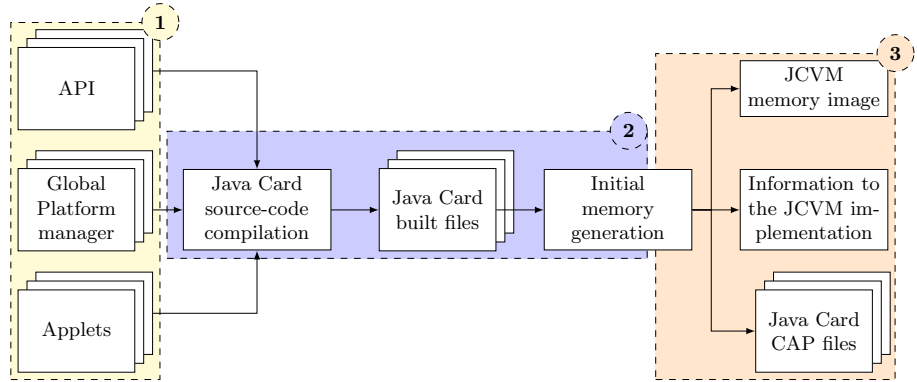
\begin{figure}[hbt]
  \centering
  \resizebox{\linewidth}{!}{
  \begin{tikzpicture}
    \matrix (m) [matrix of nodes, ampersand replacement=\&, column sep =0.75cm, row sep = 0.6cm,
    align=center,
    nodes = {
      anchor=center,
      text width = 6.5em,
      draw,
      fill=white,
    },
    doc/.style={
      draw, minimum height=4em, minimum width=3em, text width=5em,
      fill=white,
      % double copy shadow={shadow xshift=4pt, shadow yshift=4pt, fill=white, draw}
      double copy shadow={shadow xshift=4pt, shadow yshift=4pt, draw}
    },
    ]
    {
      |[doc]| {\acs{API}} \& \& \&
      \& \ac{JCVM} memory image
      \\
      |[doc]| {Global Platform manager}
      \& Java Card source-code compilation
      \& |[doc]|  {Java Card built files}
      \&  Initial memory  generation
      \& Information to the \acs{JCVM} implementation
      \\
      |[doc]| {Applets} \& \& \&
      \& |[doc]|  {Java Card \acs{CAP} files}
      \\
    };

    \draw [-latex] ($(m-1-1.east)+(8pt,8pt)$) -| (m-2-2);
    \draw [-latex] ($(m-2-1.east)+(8pt,0pt)$) -- (m-2-2);
    \draw [-latex] ($(m-3-1.east)+(8pt,8pt)$) -| (m-2-2);

    \draw [-latex] (m-2-2) -- (m-2-3);

    \draw [-latex] (m-2-3) -- (m-2-4);

    \draw [-latex] (m-2-4) -| ($(m-2-4.east)+(0.3,0)$) |- (m-1-5);
    \draw [-latex] (m-2-4) -- (m-2-5);
    \draw [-latex] (m-2-4) -| ($(m-2-4.east)+(0.3,0)$) |- (m-3-5);

    \begin{scope}[on background layer]
      \node (fit_1) [dashed, draw,
      fit={($(m-1-1.north east)+(8pt,8pt)$)(m-3-1)}, fill=yellow!20]{};

      \node (fit_2)  [dashed, draw,
      fit={($(m-2-2.west)+(-4.0pt,0)$)(m-2-3)($(m-2-4.north)+(10.0pt,14pt)$)}, fill=blue!20]{};

      \node (fit_3)  [dashed, draw, fit={($(m-2-4.north)+(20.0pt,0)$)(m-1-5)(m-2-5)(m-3-5)}, fill=orange!20]{};
    \end{scope}

    \node[circle, dashed, draw, fill=yellow!20] at (fit_1.north east) {\textbf{1}};
    \node[circle, dashed, draw, anchor=east, fill=blue!20] at (fit_2.north east) {\textbf{2}};
    \node[circle, dashed, draw, fill=orange!20] at (fit_3.north east) {\textbf{3}};

  \end{tikzpicture}
  }
  \caption{Java Card initial memory generation overview.}
  \label{fig:rommask_generation}
\end{figure}

Based on public information, documentations and literature, this article targets
how to generate \ac{JCVM} memory and how to organize it regarding Java Card
expectations. Figure~\ref{fig:rommask_generation} summarizes each step required
to generate the \ac{JCVM} initial memory state and dependences with \ac{JCVM}
implementation. Our contribution corresponds to the
part~\tikz[baseline=0]{\node[circle, dashed,anchor=base, draw, scale=0.75,
  fill=orange!20] {\textbf{3}};}. Parts~\tikz[baseline=0]{\node[circle,
  dashed,anchor=base, draw, scale=0.75, fill=yellow!20] {\textbf{1}};}
and~\tikz[baseline=0]{\node[circle, dashed,anchor=base, draw, scale=0.75,
  fill=blue!20] {\textbf{2}};} correspond to an extraction of the Java Card
specifications~\cite{book/oracle15/jcvm_spec,book/oracle15/jcre_spec,book/oracle15/api_spec,book/oracle15/dev_kit}
and GlobalPlatform specification~\cite{book/spec/gp_spe11}. Therefore, our
contribution includes:

\begin{enumerate}
  \item Which information should be extracted from every Java Card source code?
        Some information are the outputs of the Java Card toolchain. However,
        in case of native methods in Java source code, the Java Card toolchain
        cannot generate binary files. Indeed, the Java Card specification does
        not define how to handle Java native code~\cite[section
        3.3]{book/oracle15/jcvm_spec}. Therefore, in every case, we described
        how to generate binary information to put in memory and how to organize inside them.        \label{contribution:cas_1}
  \item Based on results of case~\ref{contribution:cas_1}, we introduce how
        to organize the \ac{JCVM} memory. We designed a filesystem adapted to Flash
        memory. This filesystem is designed to support the Java Card specificities
        where Java Card code and data are stored.
\end{enumerate}

Figure~\ref{fig:rommask_generation} highlights the three parts that describes article's
outline.
\begin{itemize}
  \item Part~\tikz[baseline=0]{\node[circle, dashed, anchor=base, draw,
        scale=0.75, fill=yellow!20] {\textbf{1}};} lists requirements to execute
        \ac{JCVM} and applets. Based on the Java source code,
        section~\ref{sec:platform-api} introduces the Java Card \ac{API},
        GlobalPlatform manager and applets code architecture.

  \item Part~\tikz[baseline=0]{\node[circle, dashed, anchor=base, draw,
        scale=0.75, fill=blue!20] {\textbf{2}};} introduces the compilation
        process. Since inputs introduced in part~\tikz[baseline=0]{\node[circle,
        dashed, anchor=base, draw, scale=0.75, fill=yellow!20] {\textbf{1}};} are
        built, required information should be extracted and put in \ac{JCVM}
        memory. This part is presented in the
        section~\ref{sec:building-acapi-rom}.

  \item Part~\tikz[baseline=0]{\node[circle, dashed,anchor=base, draw,
        scale=0.75, fill=orange!20] {\textbf{3}};} is the article contribution.
        Based on the
        part~\tikz[baseline=0]{\node[circle, dashed, draw, scale=0.75,
          anchor=base, fill=blue!20] {\textbf{2}};} results, we define three outputs:

        \begin{itemize}
          \item the \ac{JCVM} memory image. This image aims at setting up the
                \ac{JCVM} Flash memory. This image contains required elements
                (\acp{API}, GlobalPlatform manager, applets and other data) to
                run the Java Card platform;
          \item information requires by the \ac{JCVM} to use the memory image.
                This information is a representation of elements stored in the
                memory image. This output should be linked to the \ac{JCVM}
                source code during the compilation process. For instance, this
                information includes the list of native function tokens provided
                by the \ac{API};
          \item and a way to check if the generated elements in the \ac{JCVM} memory
                image are compliant with the Java Card security rules. This
                verification is done by the \ac{BCV}.
        \end{itemize}

        Section~\ref{sec:rom-mask-structure} describes the information extracted
        from the Java Card built files. Section~\ref{sec:flash-block} explains
        how to organize this information in memory upon a filesystem. Section~\ref{sec:experimental-results} presents our design on a ARM platform.
        Section~\ref{sec:concl-future-works} concludes this article with future
        works.

\end{itemize}

\section{\ac{JCVM} and Applets Requirements}
\label{sec:java-card-global-api}

The first step of our work is to design the \ac{JCVM} requirements to run. From
the figure~\ref{fig:rommask_generation}, it is
parts~\tikz[baseline=0]{\node[circle, dashed, draw, scale=0.75, anchor=base,
  fill=yellow!20] {\textbf{1}};} and~\tikz[baseline=0]{\node[circle, dashed,
  draw, scale=0.75, anchor=base, fill=blue!20] {\textbf{2}};}, that includes:

\begin{itemize}
  \item the \ac{API} required to execute each Java Card applets correctly,
  \item the GlobalPlatform manager
  \item and few applets can be included in the target's \ac{NVM}. These applets
        may also be installed over the air through the platform installer if
        available.
\end{itemize}

In this section, we focus on how the Java Card source code is organized and how to
build and to extract binary information required to generate initial memory.

\subsection{Java Card Platform requirements}
\label{sec:platform-api}

The Java Card platform is composed of the Java Card \ac{API}, specified by
Oracle~\cite{book/oracle15/api_spec}, and the GlobalPlatform
one~\cite{book/spec/gp_spe11}. To rely on the security provided by the Java Card
platform, this \ac{API} should be developed in Java-language. This section
introduces the part~\tikz[baseline=0]{\node[circle, dashed, draw, scale=0.75,
  anchor=base, fill=yellow!20] {\textbf{1}};} of
figure~\ref{fig:rommask_generation}.

\subsubsection{Java Card \acs{API}}

The Java Card \ac{API} is a set of functions required to execute Java Card
applets. The Java Card \ac{API} is split into three packages: \texttt{java},
\texttt{javacard} and \texttt{javacardx}. First, the \texttt{java} package
contains the Java required classes in order to run Java language baseline as
\texttt{Exception} or remote methods. Next, the \texttt{javacard} package
exposes the Java Card specific features as \ac{APDU} access or biometry
functions. Finally, the \texttt{javacardx} package contains the Java Card
features subject to export restrictions.

\subsubsection{GlobalPlatform manager}

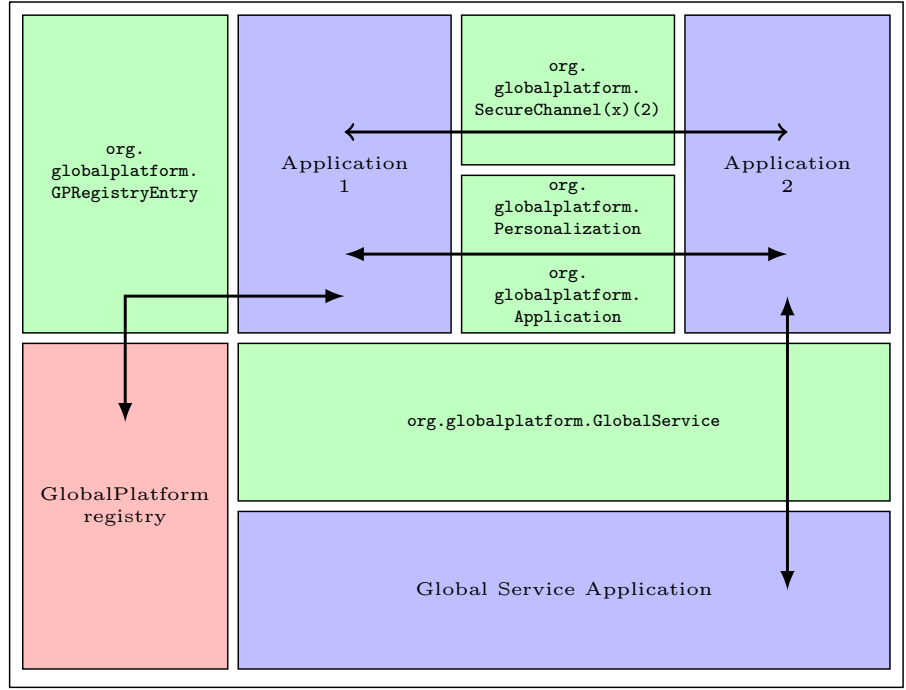
\begin{figure}[hbt]
  \centering
  \resizebox{\linewidth}{!}{
  \begin{tikzpicture}
    \def\hautG{8}; % Hauteur Globale
    \def\largG{8.5}; % Largeur Globale
    \def\quartHUnG{\largG/4-0.05};
    \def\quartHUnD{\largG/4+0.05};
    \def\quartHDeuxG{\largG/2-0.05};
    \def\quartHDeuxD{\largG/2+0.05};
    \def\quartHTroisG{\largG*0.75-0.05};
    \def\quartHTroisD{\largG*0.75+0.05};
    \def\sixVUnG{\hautG/5-0.05};
    \def\sixVUnD{\hautG/5+0.05};
    \def\sixVDeuxG{(\hautG/5)*2-0.05};
    \def\sixVDeuxD{(\hautG/5)*2+0.05};
    \def\sixVTroisG{(\hautG/5)*3-0.05};
    \def\sixVTroisD{(\hautG/5)*3+0.05};
    \def\sixVQuatreG{(\hautG/5)*4-0.05};
    \def\sixVQuatreD{(\hautG/5)*4+0.05};
    \def\larg1{\haug};

    \draw[black] (0,{\sixVUnG - 0.075}) rectangle (\largG,\hautG); % Rectangle d'ensemble
    \draw[black, fill=red!25] (0.125, \sixVUnD) rectangle (\quartHUnG, {\sixVTroisG}); % GlobalPlatform Registry
    \draw[black, fill=green!25] (0.125, {\sixVTroisD}) rectangle (\quartHUnG, \hautG-.125); % GPRegistryEntry
    \draw[black, fill=blue!25] ( \quartHUnD, {\sixVUnD}) rectangle (\largG-0.125, {\sixVDeuxG}); % Global Service Application
    \draw[black, fill=green!25] ( \quartHUnD, {\sixVDeuxD}) rectangle (\largG-0.125, {\sixVTroisG}); % GlobalService
    \draw[black, fill=blue!25] ( \quartHUnD, {\sixVTroisD}) rectangle (\quartHDeuxG, {\hautG - 0.125}); % App 1
    \draw[black, fill=green!25] ( \quartHDeuxD, {\sixVTroisD}) rectangle (\quartHTroisG, {\sixVQuatreG}); % Personalization et Application
    \draw[black, fill=green!25] ( \quartHDeuxD, {\sixVQuatreD}) rectangle (\quartHTroisG, {\hautG - 0.125}); % SecureChannel(x)(2)
    \draw[black, fill=blue!25] ( \quartHTroisD, {\sixVTroisD}) rectangle (\largG-0.125, {\hautG - 0.125}); % App 2

    % \node [font=\tiny, align=center](GPSystem) at ({\largG/2},{(\sixVUnG + 0.125)/2}){\texttt{org.globalplatform.GPSystem}};
    \node [font=\tiny, align=center](GlobalServiceApplication) at ({(\quartHUnD+(\largG - 0.125))/2},{(\sixVUnD + \sixVDeuxG)/2}){Global Service Application};
    \node [font=\tiny, align=center](GlobalService) at ({(\quartHUnD+(\largG - 0.125))/2},{(\sixVDeuxD + \sixVTroisG)/2}){\texttt{org.globalplatform.GlobalService}};
    \node [font=\tiny, align=center](App1) at ({(\quartHUnD + \quartHDeuxG)/2},{(\sixVTroisD + (\hautG-0.125))/2}){Application\\1};
    \node [font=\tiny, align=center](App2) at ({(\quartHTroisD + (\largG - 0.125))/2},{(\sixVTroisD + (\hautG-0.125))/2}){Application\\2};
    \node [font=\tiny, align=center](Perso) at ({(\quartHDeuxD + \quartHTroisG)/2},{(\sixVTroisD + \sixVQuatreG)/2}){\texttt{org.}\\\texttt{globalplatform.}\\\texttt{Personalization}\\\\\texttt{org.}\\\texttt{globalplatform.}\\\texttt{Application}};
    \node [font=\tiny, align=center](SecureChannel) at ({(\quartHDeuxD + \quartHTroisG)/2},{(\sixVQuatreD + (\hautG-0.125))/2}){\texttt{org.}\\\texttt{globalplatform.}\\\texttt{SecureChannel(x)(2)}};
    \node [font=\tiny, align=center](registryEntry) at ({(0.125 + \quartHUnG)/2},{(\sixVTroisD + (\hautG - 0.125))/2}){\texttt{org.}\\\texttt{globalplatform.}\\\texttt{GPRegistryEntry}};
    \node [font=\tiny, align=center](registry) at ({(0.125 + \quartHUnG)/2},{(\sixVUnD + (\sixVTroisG))/2}){GlobalPlatform\\registry};

    \draw[latex-latex, thick] ({(\quartHUnD + \quartHDeuxG)/2},{(\sixVTroisD + \sixVQuatreG)/2}) -- ({(\quartHTroisD + (\largG-0.125))/2},{(\sixVTroisD + \sixVQuatreG)/2});
    \draw[<->, thick] ({(\quartHUnD + \quartHDeuxG)/2},{((\sixVQuatreD + (\hautG-0.125))/2)-(\hautG/20)}) -- ({(\quartHTroisD + (\largG-0.125))/2},{((\sixVQuatreD + (\hautG-0.125))/2)-(\hautG/20)});
    \draw[latex-latex, thick] ({(\quartHTroisD + (\largG-0.125))/2},{((\sixVTroisD + \sixVQuatreG)/2)-(\hautG/20)}) -- ({(\quartHTroisD + (\largG-0.125))/2},{(\sixVUnD + \sixVDeuxG)/2});
    \draw[latex-latex, thick] ({(\quartHUnD + \quartHDeuxG)/2},{((\sixVTroisD + \sixVQuatreG)/2)-(\hautG/20)}) -- ({(0.125 + \quartHUnG)/2},{((\sixVTroisD + \sixVQuatreG)/2)-(\hautG/20)}) -- ({(0.125 + \quartHUnG)/2},{(\sixVDeuxD + \sixVTroisG)/2});

  \end{tikzpicture}
  }
  \caption{Depiction of an accurate use of the GlobalPlatform \ac{API} at runtime. Applications are represented in blue color, interfacing tools are in green, while the GlobalPlatform registry, as a database component, is red.}
  \label{fig:GP_use}
\end{figure}

A GlobalPlatform-compatible security manager must provide an \ac{API} capable of
processing various types of requests. This \ac{API} has to organize its
functions and interfaces into several parts. Figure~\ref{fig:GP_use} represents
some of these components in a correct, specification-compliant structure. All
the \ac{API}'s content must be organized under the \texttt{org.globalplatform}
package.

The framework is composed solely of interfaces, except for one class,
\texttt{GPSystem}. It is frequently a starting point for third-party platform
contents as it provides required instances to benefit from various features. It
also allows an application to query information about itself or the platform in
general, including life-cycle states.

\texttt{GlobalService} is an interface needed to query a service from an
application offering global services. An instance of it must first be obtained
from \texttt{GPSystem}, and then used to claim a \ac{SIO}, which, itself, can
provide the desired actions. While the \texttt{GlobalService} interface's
content is known and specified, the content of the \ac{SIO} is up to the global
service application, and the receiver is expected to be aware of it.

In order to get information from the GlobalPlatform registry or even modify an
entry, a third-party program must get an instance of the
\texttt{GPRegistryEntry} interface from the \texttt{GPSystem} class. It will
then be able to query data from the registry (such as its own life cycle state
or the platform's one, its own \ac{AID}), grant privileges to it or other
applets, register services or relationships with other embedded applications.
Modifying entries can be possible depending on the type of affected data and the
permission granted to the claimant.

Secure channels are an important feature designed by the GlobalPlatform
specification. It allows applications to delegate security measures for
communications, such as encryption or decryption, authentication, and security
levelling. Therefore, the developer can rely on the trusted entity to take
responsibility for all these aspects. An instance of the \texttt{SecureChannel}
interface, or its two extensions \texttt{SecureChannelx} and
\texttt{SecureChannelx2} can be obtained \textit{via} \texttt{GPSystem}.

In case cross-application communication is needed, the \texttt{Application} and
\texttt{Personalization} interfaces provide a mechanism where the implemented
applet can simply receive data. The latter also allows outputting a response. In
contrast to what the names might suggest, it can be used for any type of
message.

As described in \cite{book/spec/etsi_ts_102_226}, an \ac{HTTP} session can be set
up to simplify administration. An implementation of \texttt{HTTPAdministration}
can ask for initialization and get notified with the end result if
\texttt{HTTPReportListener} is also implemented.

A platform owner verification is usually required before conducting a sensitive operation.
A \ac{CVM} is generally provided as a global service through a \texttt{CVM}
interface.

Finally, \texttt{Authority} and \texttt{AuthoritySignature} are interfaces that,
once implemented, result in global services performing data signing and key
recovery at the request of applications.

\subsection{Building Java Card Source Code}\label{sec:building-acapi-rom}

After being developed, the Java Card \ac{API}, GlobalPlatform manager and
applets have to be built to be included into the \ac{JCVM} memory. This step
matches to figure~\ref{fig:rommask_generation}
part~\tikz[baseline=0]{\node[circle, dashed, draw, scale=0.75, anchor=base,
  fill=blue!20] {\textbf{2}};} and aims at creating the initial state
requirements used by the \ac{JCVM} implementation.

\subsubsection{Building \ac{API} from Scratch}\label{sec:building-api-from}

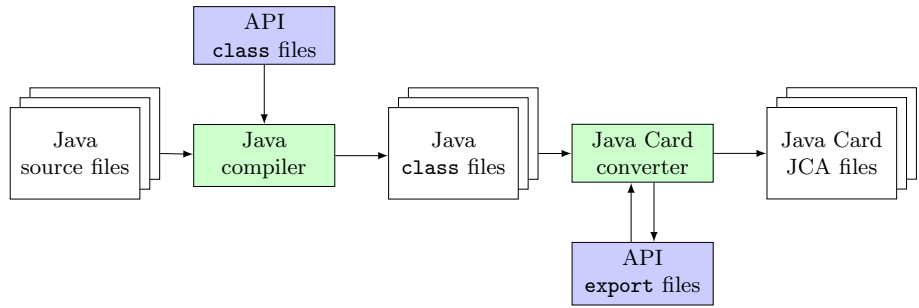
\begin{figure}[!hbt]
  \centering
  \resizebox{\linewidth}{!}{
  \begin{tikzpicture}[
    auto,
    base/.style={
      rectangle,
      draw,
      text width = 5.5em,
      align = center,
    },
    frame/.style={
      base,
      fill=blue!20,
    },
    conver/.style={
      base,
      fill=green!20,
    },
    doc/.style={
      draw, minimum height=4em, minimum width=3em, text width=5em,
      fill=white,
      % double copy shadow={shadow xshift=4pt, shadow yshift=4pt, fill=white, draw}
      double copy shadow={shadow xshift=4pt, shadow yshift=4pt, draw}
    },
    ]

    \matrix [
    matrix of nodes,
    nodes in empty cells,
    ampersand replacement=\&, column sep=0.75cm,row sep=0.6cm, align=center] (m) {%
      \&
      |[frame]| {\ac{API} \texttt{class} files}
      \\
      |[doc]| {Java source files}
      \& |[conver]| {Java compiler}
      \& |[doc]| {Java \texttt{class} files}
      \& |[conver]| {Java Card\\converter}
      \& |[doc]| {Java Card \ac{JCA} files}
      \\
      \& \& \& |[frame]| {\ac{API} \texttt{export} files}
      \\
    };

    % Arrows
    \draw [-latex] ($(m-2-1.east)+(8pt,0pt)$ -| m-2-2) -- (m-2-2);
    \draw [-latex] (m-2-2) -- (m-2-2 -| m-2-3.west);
    \draw [-latex] (m-1-2) -- (m-2-2);

    \draw [-latex] ($(m-2-3.east)+(8pt,0pt)$ -| m-2-4) -- (m-2-4);
    \draw [-latex] (m-2-4) -- (m-2-4 -| m-2-5.west);

    \draw [-latex] ($(m-2-4.south)+(0.5em,0)$) -- ($(m-3-4.north)+(0.5em,0)$) ;
    \draw [latex-] ($(m-2-4.south)-(0.5em,0)$) -- ($(m-3-4.north)-(0.5em,0)$) ;
  \end{tikzpicture}
  }
  \caption{Java Card building process from Java source files.}
  \label{fig:jca-building}
\end{figure}

The Java Card building process is summarized in figure~\ref{fig:jca-building}.
When an applet developer builds its code (its own \ac{API} or applets), he must
use the \ac{API} provided information for the target to link his code. For
instance, Oracle provides Java Card \ac{API} \texttt{class}\footnote{These
  \texttt{class} files only contain the method signatures.} and \texttt{export}
files in the Java Card SDK.\ It is the \ac{API} image used by the compiler to
build Java \texttt{class} files and, according to the \texttt{export}
files, the \ac{JCA} files.

In this work, we aim at having a reproducible compilation process with minimal
external dependencies. Decreasing the external dependencies reduces the risk to
import buggy code. However, the code to build should be analysed to avoid
security issues.

Most of the Java Card \ac{API} packages are dependent between them. In order to
find the best compilation order, we analyse their dependencies. It turns out
that the Java Card \ac{API} must be built in this order:
\begin{multicols}{2}
\begin{enumerate}
  \item \texttt{java.lang}
  \item \texttt{java.io}
  \item \texttt{java.rmi}
  \item \texttt{javacard.framework}
  \item \texttt{javacard.framework.service}
  \item \texttt{javacard.security}
  \item \texttt{javacardx.crypto}
  \item \texttt{javacardx.biometry}
  \item \texttt{javacardx.external}
  \item \texttt{javacardx.biometry1toN}
  \item \texttt{javacardx.security}
  \item \texttt{javacardx.framework.util}
  \item \texttt{javacardx.framework.math}
  \item \texttt{javacardx.framework.tlv}
  \item \texttt{javacardx.framework.string}
  \item \texttt{javacardx.apdu}
  \item \texttt{javacardx.apdu.util}
  % \item \texttt{org.globalplatform}
\end{enumerate}
\end{multicols}

The \texttt{export} files need building just after compiling the \texttt{class}
files. The Java Card converter next translates \texttt{class} files to Java Card
file using information contained in the \texttt{export} files. As indicated in
section~\ref{sec:java-card-practice}, the Java Card converter outputs \ac{CAP}
by default. However, as native methods declaration may have been declared in the
Java Card \ac{API}, \ac{CAP} files cannot be built. Native methods are not
supported in the \ac{CAP} file~\cite{book/oracle15/jcvm_spec}. Therefore, an
intermediate file format must be used: the \ac{JCA} files. So, for each \ac{API}
and GlobalPlatform package, a \ac{JCA} file is obtained.

\subsubsection{Parsing the \ac{JCA} Files}\label{sec:parsing-jca-file}

Now that the pool of \ac{JCA} files is obtained, it should be parsed to extract
required information to generate the \ac{JCVM} initial memory.\newline

The \ac{JCA} file is a text file generated by the Java Card converter as a human
representation of the \ac{CAP} file. This file is not executable as is.

A syntax analyser must be developed to exploit the information contained in this
file. The \ac{JCA} file can be compiled into a \ac{CAP} file using the Java Card
converter.

Oracle publicly and partially defines the \ac{JCA} file in the Java Card
development kit~\cite{book/oracle15/dev_kit} through a simple example. As
introduced in the development kit, the \ac{JCA} file contains the following
information for a Java Card package:

\begin{itemize}
  \item imported packages: the package \ac{ID} required by the current one;
  \item declared applets references;
  \item constant pool with type of each element used in each class's methods;
  \item classes and interfaces associated with the current package. For each class
    or interface, we have:

  \begin{itemize}
    \item fields declaration;
    \item virtual methods tables;
    \item interfaces table;
    \item remote interfaces table;
    \item implemented methods with their bytecodes.
  \end{itemize}
\end{itemize}

To parse the \ac{JCA} file, we propose to use the \texttt{javacc}
lexer~\cite{journals/spe/GuptaN16}. This tool, developed in Java, generates,
from a \ac{BNF} grammar, a Java-parser library. So, to parse the \ac{JCA} file,
the grammar must be defined. In the Oracle's development
kit~\cite{book/oracle15/dev_kit}, Oracle points out that this grammar is
available in the ``\textit{source release}'' version. As this version is not
public, we do not have access to it. \newline

To write the \ac{BNF} grammar of the \ac{JCA} file, we have played with the
compilation parameters and the program to build. We analyse how the Java method
accessors (\texttt{native}, \texttt{final}, \textit{etc.}) and tokens are
organized and so on. Therefore, we developed a set of Java Card application
codes where each Java keywords are used. Next, we analyse the obtained \ac{JCA}
files to understand the relationship between each keyword and its representation
in the \ac{JCA} file. Listing~\ref{fig:overview-jca-file} gives the big picture
of the \ac{JCA} file grammar.

\begin{lstlisting}[caption={An overview of the \ac{JCA} file.}, label={fig:overview-jca-file}]
.package PACKAGE_NAME {
  .aid PACKAGE_ID
  .version MAJOR.MINOR
  .imports { [LIST OF (PACKAGE_ID VERSION_MAJOR.MINOR)] }
  .applets { [LIST OF (CLASS_ID classname)] }

  .constant pool
    { [ENTRIES OUT OF CURRENT PACKAGE] [ENTRIES IN THE CURRENT PACKAGE] }

  .class [ACCESSOR] [abstract] [final]
                    classname [extends (NAME | TOKEN)]
    { [IMPLEMENTED SHAREABLE INTERFACES]
      [IMPLEMENTED REMOTE INTERFACES]
      [FIELDS LIST]
      [PUBLIC & PACKAGE METHOD TABLES]
      [IMPLEMENTED INTERFACE METHODS LIST]
      [IMPLEMENTED REMOTE METHOD INTERFACE METHODS LIST]
      .method [ACCESSOR] [abstract] [static] [final] [native]
        METHOD_SIGNATURE [token]
          { METHOD HEADER [METHOD BYTECODES]
                          [METHOD EXCEPTION HANDLER] }
} }
\end{lstlisting}

To understand this grammar, let's us explain how to read it:
\begin{itemize}
  \item words in uppercase signify a set of different information hidden for better reading;
  \item words in lowercase define the constant string value used as-is in the
    \ac{JCA} file;
  \item Sentences between square brackets point optional elements;
  \item Sentences between parentheses refer to mandatory element;
  \item The \texttt{|} indicates a logical OR between the occurrence of each
    element as (A\texttt{|}B) for (A or B).
\end{itemize}

Since this grammar is given to \texttt{javacc}, we obtain a Java library which
allows us to parse \ac{JCA} files.\newline

\section{Extracting Information from Java Card \ac{API}}\label{sec:rom-mask-structure}

Since we can parse \ac{JCA} files, we should analyse it to generate the initial
data. This section aims at introducing the figure~\ref{fig:rommask_generation}
part \tikz[baseline=0]{\node[circle, dashed, draw, scale=0.75, anchor=base,
  fill=orange!20] {\textbf{3}};} to define efficient initial state
outputs.\newline

The initial memory generator has three outputs:

\begin{itemize}
  \item Information to the \ac{JCVM} implementation given as a C-header file,
        named \texttt{jni.h}. This file contains a generated native method
        dispatcher, declaration of native methods to implement and the triple
        (package \ac{ID}, class \ac{ID}, method \ac{ID}) to call the starter. A
        part of this file is described in section~\ref{sec:from-java-native};

  \item The Java Card \ac{CAP} files. This output offers a way to verify
        generated \ac{CAP} files with the Java Card \ac{BCV}. The initial state
        generator fills the \ac{CAP} file structures from the parsed \ac{JCA}
        files. To ensure the correctness of this translation, \ac{CAP} files are
        generated to verify them as explained in section~\ref{sec:capmap}.

  \item The \ac{JCVM} initial state memory image via a binary file used to set
        the target's \ac{NVM}. This output is introduced in
        section~\ref{sec:flash-block};
\end{itemize}

A \ac{JCA} file is a comprehensible human version of the \ac{CAP} file and
contains the same information. When building a binary version of all the JCA
files, the simple solution is to translate each of them to a \ac{CAP} file. As
the \ac{CAP} file is executable as-is by the \ac{JCVM}, this translation may be
enough. However, some elements such as Java native method cannot be represented
in the \ac{CAP} file.

\subsection{Native Methods Call}\label{sec:from-java-native}

The Java Card specification does not define any way to call the native method.
However, it defines two bytecodes, \texttt{impdep1} and \texttt{impdep2}, as:
\begin{quotation}
  ``\textit{The two reserved opcodes, numbers 254 and 255, have the mnemonics
    \texttt{impdep1} and \texttt{impdep2}, respectively. These instructions are
    intended to provide ``back doors'' or traps to implementation-specific
    functionality implemented in software and hardware, respectively.}''
  \begin{flushright}
    (\ac{JCVM} specification~\cite[Section~7.2]{book/oracle15/jcvm_spec})
  \end{flushright}
\end{quotation}

\texttt{impdep1} and \texttt{impdep2} instructions offer a standard way to
execute a piece of code outside the \ac{JCVM} sandbox. To the best of our
knowledge, none of \ac{JCVM} developers use these instructions on their \ac{VM}
implementation. Thus, we choose these instructions to be a part of the mechanism
to call native code from the \ac{JCVM}. \newline

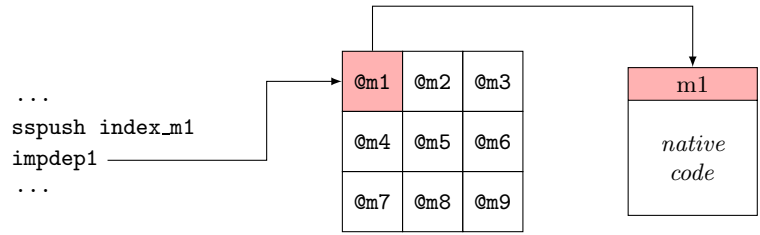
\begin{figure}[!hbt]
  \centering
  \begin{tikzpicture}[
      box/.style={
        minimum width=0.8cm,
        minimum height=0.8cm,
      },
      rec/.style={
        draw,
        rectangle split,
        rectangle split parts=2,
        text centered,
        rectangle split part fill={red!30, white},
    },
      nodes={
        box,
      },
      row sep=-\pgflinewidth,
      column sep=-\pgflinewidth
    ]
    \matrix [matrix of nodes,
    nodes in empty cells,
    nodes={draw, anchor=center}] (native_array) {
      |[fill=red!30]|\texttt{@m1} & \texttt{@m2} & \texttt{@m3} \\
      \texttt{@m4} & \texttt{@m5} & \texttt{@m6} \\
      \texttt{@m7} & \texttt{@m8} & \texttt{@m9} \\
    };

    \node (impdep) at ($(native_array.west)-(3.0, 0.25)$) [anchor=east,box] {\texttt{impdep1}};
    \node (push) at (impdep.north west) [anchor=west,box] {\texttt{sspush index\_m1}};
    \node at (push.north west) [anchor=west,box] {\texttt{\dots}};
    \node (push) at (impdep.south west) [anchor=west,box] {\texttt{\dots}};

    \node (native) at ($(native_array.east)+(3.0, 0)$) [anchor=east,rec] {
      \nodepart[text width=1.5cm]{one} m1
      \nodepart{two} \parbox[c][\dimexpr 4em][c]{1.5cm}{\centering \textit{native code}}
    };

    % \node [draw, text width=3em, align=center] at (native.one south |- native_array.2-3) {
    %   native code};

    \draw [-latex] (impdep)
           -| ($(impdep.east -| native_array-2-1.west)-(1.0,0)$)
           |- (native_array-1-1.west);

    % \path [bend left, -latex] (native_array-1-1.north) edge (native.one north);
    \draw [-latex] (native_array-1-1) |- ($(native_array-1-1)+(0.0,1.0)$) -| (native.one north);
  \end{tikzpicture}
  \caption{Calling native code.}
  \label{fig:calling-native-code}
\end{figure}

Therefore, in order to call native code, an offset of the native method to
execute should be pushed on the Java Card stack before executing the
\texttt{impdep} instruction. As the Java Card stack contains 2-byte words, we
can thus encode native method offset on 2-byte without footprint overweight.
Therefore, we choose to push a 2-byte offset to have enough index. With 2-byte
offset, we can call up to \numprint{65535} native methods. The associated native
methods will be called through a switch-case statement.
Figure~\ref{fig:calling-native-code} gives a global overview of the process we
have developed. If an unused token is called, the \texttt{default} statement
will be executed, and a security exception will be thrown.

For this behaviour to be obtained, a dispatcher method named \texttt{callJCNativeMethod}
is generated. It requires as parameters the current Java Card context and an
index to the native method to call. The index is arbitrarily determined by the
initial state generator where the first native method found by the parser has the
index 1 and so on. We rely on the \ac{JCVM} implementation to check whether the
current Java Card context is allowed to execute the associated native method.

As an example, let's consider the Java native
\texttt{myNativeMethod} method in listing~\ref{lst:java_native}. This method is
called by \texttt{executeNative} method which has two-byte parameters.

\begin{lstlisting}[
  language=java,
  caption={Calling native method from Java point of view.},
  label={lst:java_native}
]
class MyClass {
  native short myNativeMethod(byte p1, byte p2);

  void executeNative() {
    byte p1 = 0xCA, p2 = 0xFE;
    short ret = myNativeMethod(p1, p2);
} }
\end{lstlisting}

After building \texttt{MyClass}, the \texttt{jni.h} file is generated and
contains the \ac{JCVM} to native methods dispatcher and native methods signature
to implement. A part of the \texttt{jni.h} file is shown in the
listing~\ref{lst:dispatcher}.

\begin{lstlisting}[
  language=c++,
  numbers=left,
  caption={Native method dispatcher and native method to implement in \texttt{jni.h}.},
  label={lst:dispatcher}]
extern jshort_t MyClass_myNativeMethod (jbyte_t p1,
                                        byte_t p2);
// ...
void callJCNativeMethod(Context& context, jshort_t index) {
  Stack& stack = context.getStack();
  Heap& heap = context.getHeap();
  switch(index) {
    case MYCLASS_MYNATIVEMETHOD: {
      jbyte_t p2 = stack.pop_Byte();(*\label{lst:dispatcher:begin}*)
      jbyte_t p1 = stack.pop_Byte();(*\label{lst:dispatcher:begin1}*)
      jshort_t ret = MyClass_myNativeMethod(p1, p2);(*\label{lst:dispatcher:call}*)
      stack.push_Short(ret);(*\label{lst:dispatcher:end}*)
      break;
    // ...
    default:
      // If the native method index is unknown, a Java Security Exception must be thrown!
      throw SecurityException();
} } }
\end{lstlisting}

In listing~\ref{lst:dispatcher}, the \texttt{myNativeMethod} method execution is
made from line~\ref{lst:dispatcher:begin} to \ref{lst:dispatcher:end}. The first
two lines (\ref{lst:dispatcher:begin} and \ref{lst:dispatcher:begin1}) push all
parameters from the Java Card stack to the C-stack required to execute the
native method. On line~\ref{lst:dispatcher:call}, the native method is executed
and its returned value is saved and pushed on the Java Card stack
(line~\ref{lst:dispatcher:end}). For each native Java method, our tool generates
this kind of fragment of the code required to jump from the Java Card side to
the native one. As the native method is declared as \texttt{extern}, we rely on
the \ac{JCVM} developer for implementing it. Thus, the C-compiler will link the
dispatcher calls to the native method implementations.\newline

As this piece of code is executed out of the Java Card sandbox, it should be
segregated by the \ac{OS} to protect the system if malicious code is running.
Bouffard \etal{}~\cite{conf/sstic/bouffard18} proposed to use the component
\ac{MPU} to ensure this security property.

\subsection{Creating the \ac{CAP} Files Structures}\label{sec:capmap}

As introduced in section~\ref{sec:java-card-practice}, the \ac{CAP} file is
designed to be executed as-is by the \ac{JCVM}. To be compliant with this
architecture, we parse the \ac{JCA} files to fill the \ac{CAP} structures.

\subsubsection{The \ac{CAP} File}
\label{sec:cap-file}

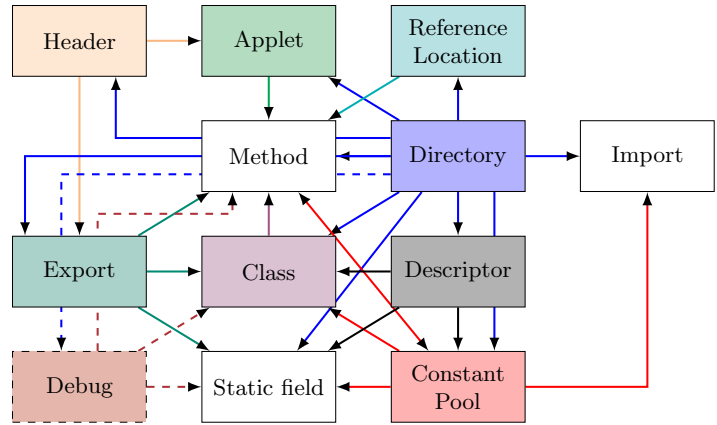
\begin{figure}[!hbt]
  \centering
  \resizebox{0.8\linewidth}{!}{
    \begin{tikzpicture}
      \matrix (m) [
      matrix of nodes,
      ampersand replacement=\&,
      column sep =0.75cm,
      row sep = 0.6cm,
      align=center,
      nodes = {
        anchor=center,
        text width = 5em,
        minimum height=3em,
        draw,
        fill=white,
      },
      ]
      {
        |(header)[fill=Apricot!30]|Header \& |(applet)[fill=Green!30]|Applet \& |(rl)[fill=TealBlue!30]|Reference Location         \&  \\
        \& |(method)|Method \& |(directory)[fill=blue!30]|Directory     \& |(import)|Import \\
        |(export)[fill=PineGreen!30]|Export \& |(class)[fill=DarkOrchid!30]|Class  \& |(descriptor)[fill=black!30]|Descriptor    \&        \\
        |(debug)[fill=Maroon!30, dashed]|Debug   \& |(static)|Static field \& |(cp)[fill=red!30]|Constant Pool \&        \\
      };

      \begin{scope}[on background layer]
        % Links from the Header
        \draw[-latex, thick, draw=Apricot] (header) -- (applet); % to Applet
        \draw[-latex, thick, draw=Apricot] (header) -- (export); % to Export

        % Links from the Directory
        \draw[-latex, thick, draw=blue] (directory) -- (rl);
        \draw[-latex, thick, draw=blue] (directory) -- (applet);
        \draw[-latex, thick, draw=blue] (directory) -- (import);
        \draw[-latex, thick, draw=blue] (directory) -- (method);
        \draw[-latex, thick, draw=blue] (directory) -- (descriptor);
        \draw[-latex, thick, draw=blue] (directory) -- (class);
        \draw[-latex, thick, draw=blue] ($(directory.south)+(0.5,0)$) -- ($(cp.north)+(0.5,0)$);
        \draw[-latex, thick, draw=blue] ($(directory.south)-(0.5,0)$) -- (static);
        \draw[-latex, thick, draw=blue] ($(directory.west)+(0,0.25)$) -| ($(header.south)+(0.5,0)$);
        \draw[-latex, thick, draw=blue] (directory.west) -| ($(export.north)-(0.75,0)$);
        \draw[-latex, thick, draw=blue, dashed] ($(directory.west)-(0,0.25)$) -| ($(debug.north)-(0.25,0)$);

        % Links from the Applet
        \draw[-latex, thick, draw=Green] (applet) -- (method);

        % Links from the Constant Pool
        \draw[-latex, thick, draw=red] (cp) -| (import);
        \draw[-latex, thick, draw=red] (cp) -- (static);
        \draw[-latex, thick, draw=red] (cp) -- (class);
        \draw[latex-latex, thick, draw=red] (cp) -- (method);

        % Links from the Class
        \draw[-latex, thick, draw=DarkOrchid] (class) -- (method);

        % Links from the Reference Location
        \draw[-latex, thick, draw=TealBlue] (rl) -- (method);

        % Links from the Export
        \draw[-latex, thick, draw=PineGreen] (export) -- (method);
        \draw[-latex, thick, draw=PineGreen] (export) -- (class);
        \draw[-latex, thick, draw=PineGreen] (export) -- (static);

        % Links from the Descriptor
        \draw[-latex, thick] (descriptor) -- (class);
        \draw[-latex, thick] (descriptor) -- (cp);
        \draw[-latex, thick] (descriptor) -- (static);

        % Links from the Debug
        \draw[-latex, thick, dashed, draw=Maroon] ($(debug.north)+(0.25,0)$) |- ($(method.south)-(0.5,0.3)$)
        -| ($(method.south)-(0.5,0)$);
        \draw[-latex, thick, dashed, draw=Maroon] (debug) -- (class);
        \draw[-latex, thick, dashed, draw=Maroon] (debug) -- (static);
      \end{scope}

    \end{tikzpicture}
  }
  \caption[\ac{CAP} components dependencies]{\ac{CAP} components dependencies. A
    component pointed is referenced by the component at the other extremity of
    the arrow. Only the Method component has a double dependency on Constant
    Pool component. A token in the Method component refers to an entry in the
    Constant Pool component which may refer to a method in the Method component.
    Dashed arrows link dependencies to or from debug component.}
  \label{fig:capfile}
\end{figure}

The \ac{CAP} file is based on components. It possesses ten standard
components: Header, Directory, Import, Applet, Class, Method, Static Field,
Export, Constant Pool, Reference Location and one optional: Descriptor.
Moreover, the targeted \ac{JCVM} may support user custom components. We exclude
the Debug component because it is only used on the debugging step and it is not
loaded in the platform. Each component is interconnected.
Figure~\ref{fig:capfile} summarizes the dependencies between each component.

\subsubsection{From the \ac{JCA} to the \ac{CAP} File}
\label{sec:from-jca-cap}

\begin{figure}[!hbt]
  \centering
  \resizebox{0.8\linewidth}{!}{
    \begin{tikzpicture}
      \matrix (m) [matrix of nodes, ampersand replacement=\&, column sep =0.60cm, row sep = 0.6cm,
      align=center,
      nodes = {
        anchor=center,
        text width = 4.5em,
        minimum height=3em,
        draw,
        fill=white,
      },
      ]
      {
        |(jca)|JCA file\\
         \& |(static)|Static Field\\
         \& |(import)|Import \&\\
         \& |(method)|Method \&  \& \\
         \& |(rl)|Reference Location \& |(applet)|Applet \&|(header)|Header \& |(export)|Export \\
         \& |(class)|Class \& \&  \& \\
         \& |(cp)|Constant Pool  \& \& \& |(descriptor)|Descriptor \&\\
      };

      \draw [-latex] (jca) |- (static);
      \draw [-latex] (jca) |- (import);
      \draw [-latex] (jca) |- (method);
      \draw [-latex] (jca) |- (class);
      \draw [-latex] (jca) |- (cp);

      \draw [-latex] (jca) |- ($(rl.south)!0.5!(class.north)$) -| (applet);
      \draw [-latex] (method) -| (applet);

      \draw [-latex] (method) -- (rl);

      \draw [-latex] (method) -| ($(export.north)-(0.25,0)$);
      \draw [-latex] (class)  -| (export);
      \draw [-latex] (static) -| ($(export.north)+(0.25,0)$);

      \draw [-latex] (class) -| (descriptor);
      \draw [-latex] (cp) -- (descriptor);

      \draw [-latex] (export) -- (header);
      \draw [-latex] (applet) -- (header);
      \draw [-latex] (class) -| (header);

      \node [] (tmp) at ($(static.north west)-(0.1,0)$) {};
      \node (myfit) [dashed, draw, fit = (tmp)(descriptor)] {};
      \node at ($(myfit.north east)+(-0.25,0)$)
      [text width = 4.5em, minimum height=3em, draw, fill=white, align=center] {Directory};

    \end{tikzpicture}
  }
  \caption[\ac{CAP} components construction order]{\ac{CAP} components
    construction order. Arrows specify information propagation from the
    \ac{JCA} file to each \ac{CAP} components.}
  \label{fig:cap_construction_order}
\end{figure}
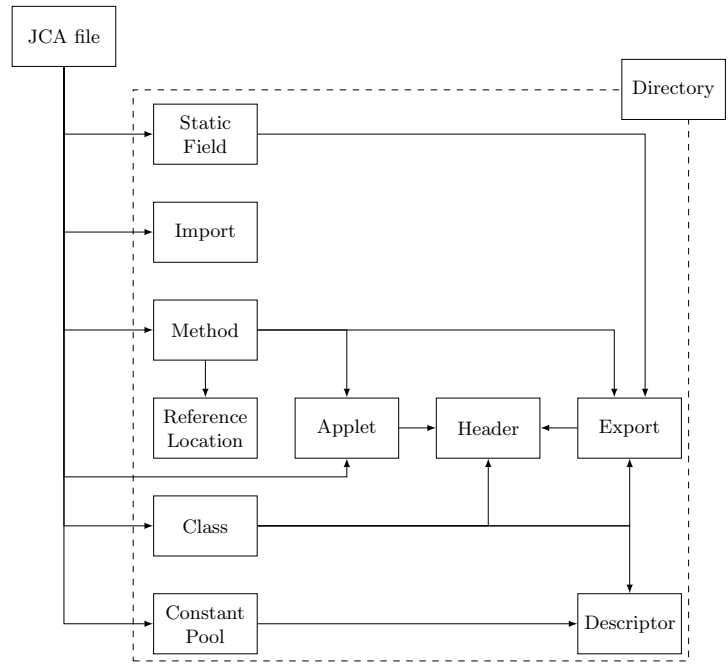

From the \ac{JCA} file grammar in listing~\ref{fig:overview-jca-file}, one sees
that this file contains partial information to build the \ac{CAP} file. This
information should be parsed and analysed to fill the \ac{CAP} component. From
the \ac{JCA} file, we can build Static Field, Import, Method, Constant Pool and
Class components. The Export component is built from the Static, Method and
Class components. The Descriptor component is constructed based on the content
of Constant Pool and Class components. Reference Location component is generated
based on the Method component. Applet component is built from the \ac{JCA} file
and the offset of the Method component. The Header component is based on Applet,
Class and Export components state. Finally, the Directory component contains the
size of all other components. Figure~\ref{fig:cap_construction_order}
synthesizes the \ac{CAP} components construction order.

To fill the \ac{CAP} structures and to ensure the coherence between components,
we base this step on the Java-library CapMap\footnote{Available on:
  \url{https://bitbucket.org/ssd/capmap-free}}.

\subsubsection{Testing the Generated \ac{CAP} Components}
\label{sec:test-gener-cap}

To verify the coherence of the generated \ac{CAP} components, each of them must
be verified by the Java Card \ac{BCV}. Included in the Java Card SDK, the
\ac{BCV} aims at checking the structure and the semantic for the input \ac{CAP}
file. An \texttt{export} files for each imported package is required to process
this analyse.\newline

In our setup, CapMap gives us \ac{CAP} files to be verified by the \ac{BCV}. The
\texttt{export} files are obtained during the compilation of the Java source
files. However, the \ac{CAP} files which contain \texttt{impdep1} or
\texttt{impdep2} bytecodes cannot be verified. Indeed, the \ac{BCV} does not
know the associated semantics of these instructions nor which values are popped and
pushed. \newline

To decrease the number of unchecked packages by the \ac{BCV}, we recommend
separating native methods in a specific package which only contains
signature of native methods to implement. Therefore, except for this package,
each other must be successfully validated by the \ac{BCV}.

\section{Java Card Initialization Memory Image}\label{sec:flash-block}

The Java Card initialization image is a binary blob which contains enough data
to set the target's \ac{NVM} and start the \ac{JCVM} instance.\ That includes at
least:

\begin{itemize}
  \item Packages table where each installed package is listed. This table
    should be used to check if a package is present or not. This table is mainly
    provided for optimization;

  \item Applications (\ac{API}, GlobalPlatform manager and applets) serialized
        data: the data to serialize were obtained as explained in the previous
        section;

  \item Non-null static fields initial values: on Java Card, static fields are
    accessible (read and write) by all contexts. Therefore, we decided to put this
    data out of the application memory area to rely on the \ac{JCVM} implementation
    for checking access rights. Indeed, application code can only be read
    whereas the static fields which can be read and written;

    % \item Applet field value.
\end{itemize}

For each of these elements to be stored, a file system has been designed. This
file system aims at providing an easy access to \ac{NVM} organization. On recent
microcontrollers, the \ac{NVM} is mainly Flash memory.

\subsection{File System Constraints on Flash Memory}
\label{sec:fs-constraints}

The Flash memories embedded in the microcontroller are divided by sectors. Each
sector might not have the same size, nonetheless it is always a power of 2.
Each Flash bit can move from 1 to 0 when a writing operation happens. However,
to shift a bit from 0 to 1, all the sector must first be erased, then set to 1.\newline

On an advanced microcontroller, Flash memory can be split in two same size
banks, divided in same number of same size sectors. The banks can be exchanged
to roll back to the previous system state when an error on a sensitive operation
occurs. For instance, the Wookey project~\cite{conf/acsac/BenadjilaMRTT19} uses
this functionality on the update system to restore the previous version when an
update fails. Flash memory banks are out of the scope of this work and we rely
on the \ac{JCVM} implementation for switching between them. Therefore, the
initial state generation aims at generating the initial state of the main Flash
bank.

\subsection{Related Works on File System for Flash Memories}
\label{sec:state-art-embedded-fs}

For efficient storage of data in Flash memory, a file system should be adapted.
A first step is to check the state-of-the-art Flash file system to find the best
ratio metadata length to data access speed.\newline

Included in the Linux kernel, several file systems designed for the Flash memory
have been developed. For instance, we can cite JFFS2~\cite{woodhouse2001jffs},
YAFFS~\cite{sun2014research}, UBIFS~\cite{conf/fm/SchierlSHR09} or
LogFS~\cite{conf/sensys/DaiNH04} developed to access to the Flash memory
directly. These file systems are the main ones used to manage the content of
Flash memories. However, they require more memory than limited-resource IoT
devices have. For this reason, we developed a small-memory footprint file system
according to the Java Card requirements.

\subsection{Design of the File System}
\label{sec:impl-fs}

The file system we designed requires a hash table, stored only in \ac{RAM}, and
rebuilt at each microcontroller starts by scanning the entire Flash memory. This
hash table is a structure contains the size and the address of each valid block
in Flash memory. This approach is possible due to its small size.

The writing operation consists in adding a block in Flash memory with the new
key and content value and updating the hashtable in \ac{RAM}. The update of the
content value will then be automatically taken into account at the next startup,
thanks to the full scan of the Flash memory. In order to make the operation
atomic, few details should be taken into account. First, a block will be written
in memory and it is tagged as an invalid data. Next, block is tagged as valid
when the block writing process is finished. Moreover, each block is assigned
a checksum to manage a card withdrawal risk in the middle of an erasure
operation, which could leave the platform in an undetermined state.

Since the only operations available are to add a new block in a sector and to
erase a whole sector, sectors will naturally fill-up. It is therefore necessary
to remove the blocks marked invalid at some point. The method chosen to carry
out this operation is to copy each block that is still valid into another
sector; as if the block had been modified, and, therefore, benefiting from the
same atomicity guarantees. Then, one erases the Flash sector that no longer
contains a valid block.

It is necessary to ensure we always have enough Flash space to perform this
``defragmentation'' operation, so a sector is reserved for this.

\subsubsection{Format of a Sector}

A sector is the concatenation of blocks. A block consists in either a
written block, as defined below, or an empty block (\textit{i.e.} not yet written:
all bits are set to 1) or of an erased block; \textit{i.e.} all the bits are set
to 0.

\subsubsection{Format of a Block}

\begin{figure}[!hbt]
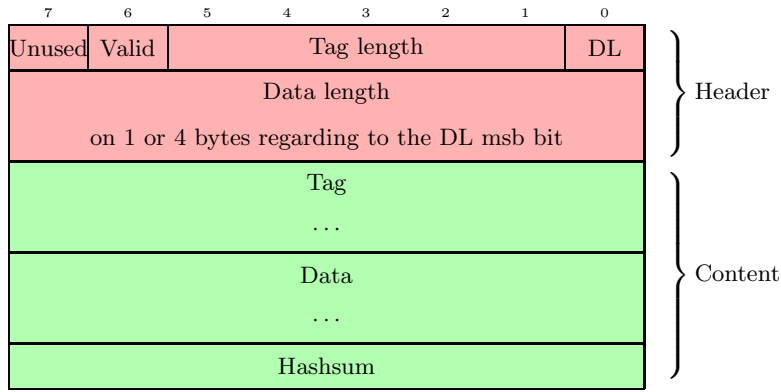

  \centering
  \begin{bytefield}[bitwidth=3.25em,endianness=big]{8}
    \bitheader{0-7} \\
    \begin{rightwordgroup}{Header}
      \colorbitbox{red!30}{1}{Unused} &
      \colorbitbox{red!30}{1}{Valid} &
      \colorbitbox{red!30}{5}{Tag length} &
      \colorbitbox{red!30}{1}{DL}\\

      \rlap{\wordbox[tlr]{1}{\color{red!30}\rule{\width}{\height}}}%
      \wordbox[tlr]{1}{Data length} \\
      \rlap{\wordbox[blr]{1}{\color{red!30}\rule{\width}{\height}}}%
      \wordbox[blr]{1}{on 1 or 4 bytes regarding to the DL msb bit}
    \end{rightwordgroup}\\

    \begin{rightwordgroup}{Content}
      \rlap{\wordbox[tlr]{1}{\color{green!30}\rule{\width}{\height}}}%
      \wordbox[tlr]{1}{Tag} \\
      \rlap{\wordbox[blr]{1}{\color{green!30}\rule{\width}{\height}}}%
      \wordbox[blr]{1}{$\cdots$} \\
      \rlap{\wordbox[tlr]{1}{\color{green!30}\rule{\width}{\height}}}%
      \wordbox[tlr]{1}{Data} \\
      \rlap{\wordbox[blr]{1}{\color{green!30}\rule{\width}{\height}}}%
      \wordbox[blr]{1}{$\cdots$} \\
      \rlap{\wordbox{1}{\color{green!30}\rule{\width}{\height}}}%
      \wordbox{1}{Hashsum}
    \end{rightwordgroup}

  \end{bytefield}
  \caption{File system block}
  \label{fig:fs:block}
\end{figure}

The block format is introduced in figure~\ref{fig:fs:block}.

Figure~\ref{fig:fs:block} shows header and block content. The header is made up
of several fields:
\begin{itemize}
  \item \textit{Unused}: bit-size field consists in adding by default to 1
    during a block writing. When this block is written, this bit is set to 0;
  \item \textit{valid}: bit-size field initially sets to 1 during a block
    writing operation, it is set to 0 when another block with the same tag is
    written;
  \item \textit{Tag length}: this 5-bit size field contains the data tag length.
    The binary values \texttt{0b11111} and \texttt{0b00000} are forbidden
    because they indicate an empty block or an erased block;
  \item \textit{DL (data length)}, bit-size field sets to 1 if the data length
    is encoded on 4 bytes, otherwise 0 when the data length is encoded on 1
    byte.
\end{itemize}

In the block content, one will find:

\begin{itemize}
  \item The \textit{data length field}: it contains the length of the data field,
  which is either 1 or 4 bytes depending on DL.\ If it is stored on 4 bytes,
  then it is stored with the most significant byte on the left;
  \item The tag and data fields;
  \item The \textit{hashsum field}: it is the result of the CRC-8 of all other
  fields except \textit{Unused} and \textit{Valid} ones.
\end{itemize}

\subsection{Using File System for Java Card Serialized Data}
\label{sec:using-fs}

\subsubsection{Global overview}

Now that we have specified the file system designed to serialize Java Card data,
we focus on how to structure data using this file system. We want it to store
a package list, \ac{CAP} structures (for \ac{API}, GlobalPlatform manager and applets),
static and applet fields. The tag field is used to encode the type of stored
data. More specifically the first byte of the tag is dedicated of this
purpose. Formally, we have for the different types:

\begin{itemize}
  \item For the package list, the tag value is (0) and its content is a binary
        array where each bit marks whether if the package is installed (bit set)
        or not (bit reset);

  \item The \ac{CAP} structures have the tag value (\texttt{1}, \texttt{package
        number}) and its content is the result of the \ac{JCA} file translation
        obtained in section~\ref{sec:capmap};

  \item The static field tag value is (\texttt{2}, \texttt{package number},
        \texttt{static field number}). Its content can be statically computed
        regarding the \ac{CAP} static field component. This component is the
        initial vector of all static fields. For static objects, the default
        value is \texttt{null} and it must be instantiated upon a class field
        owner constructor;

  \item The applets field tag value is (\texttt{3}, \texttt{package number},
        \texttt{class number}, \texttt{applet field number)}. The applet field
        content is not statically computable. Those data must be created during
        the runtime by the \texttt{Applet} class constructor. So, we rely on the
        \ac{JCVM} to serialize the field data;

  \item all other tag values are reserved for futur uses.
\end{itemize}

By encoding the tag element on 1-Byte, our solution limit the number of packages
to 256\footnote{The Java Card specification does not limit the number of
  packages}. For the class, applet field and static field number, this size is
compliant with the one defined by the Java Card
specification~\cite{book/oracle15/jcvm_spec}.\newline

When a field is stored in Flash memory, we save its type for security reasons,
to prevent type confusion. In the Flash block data part for static and applet
field, we encoded the field type in the first byte. We use the values introduced
in \autoref{lst:field_type}.

\begin{lstlisting}[language=java,
   caption={Fields type values.}, label={lst:field_type}]
byte FIELD_TYPE_BYTE    = 0;
byte FIELD_TYPE_BOOLEAN = 1;
byte FIELD_TYPE_SHORT   = 2;
byte FIELD_TYPE_INT     = 3;
byte FIELD_TYPE_OBJECT  = 4;

byte FIELD_TYPE_ARRAY_BYTE   = ((1 << 7)|FIELD_TYPE_BYTE);
byte FIELD_TYPE_ARRAY_BOOLEAN= ((1 << 7)|FIELD_TYPE_BOOLEAN);
byte FIELD_TYPE_ARRAY_SHORT  = ((1 << 7)|FIELD_TYPE_SHORT);
byte FIELD_TYPE_ARRAY_INT    = ((1 << 7)|FIELD_TYPE_INT);
byte FIELD_TYPE_ARRAY_OBJECT = ((1 << 7)|FIELD_TYPE_OBJECT);

byte FIELD_TYPE_TRANSIENT_ARRAY_BYTE
                       = ((1 << 6)|FIELD_TYPE_ARRAY_BYTE);
byte FIELD_TYPE_TRANSIENT_ARRAY_BOOLEAN
                       = ((1 << 6)|FIELD_TYPE_ARRAY_BOOLEAN);
byte FIELD_TYPE_TRANSIENT_ARRAY_SHORT
                       = ((1 << 6)|FIELD_TYPE_ARRAY_SHORT);
byte FIELD_TYPE_TRANSIENT_ARRAY_INT
                       = ((1 << 6)|FIELD_TYPE_ARRAY_INT);
byte FIELD_TYPE_TRANSIENT_ARRAY_OBJECT
                       = ((1 << 6)|FIELD_TYPE_ARRAY_OBJECT);
\end{lstlisting}
\subsubsection{Using this File System}

Since the file system is designed, see how to used it. Suppose the partial Java
Card class shows in Listing~\ref{fig:fs:example:1}. This class is a Java
Card applet, named \texttt{Example}, associated to the \texttt{sample} package.

\begin{lstlisting}[language=java,
   caption={A part of Java Card Applet to serialized.}, label={fig:fs:example:1}]
package Sample;

class Example {
   static byte[]  static_field = new byte[]
                                    {0x01, 0x02, 0x03, 0x04};
   private short  class_field1 = (short) 0x4567;
   private Object class_field2 = new Object();
   // ...
}
\end{lstlisting}

As the file system is a set of blocks parsed in Flash memory. We will describe,
in this section, which value has the blocks stored in memory regarding the
example.\newline

\texttt{Example} class will be serialized with 4 block: one for the \texttt{Example}
class representation and ones for every class field. \texttt{Example} class
and \texttt{static\_field} blocks are statically computed during the Java Card
initialization memory image generation. However, \texttt{class\_field1} and
\texttt{class\_field2} blocks are computed during the \texttt{Example} class
constructor when fields are instantiated.

By supposing the \texttt{sample} package is identified with number 3, and each
field is initialized, we will have:

\begin{enumerate}
  \item \texttt{sample} \ac{CAP} file block value, this \ac{CAP} has the
        \texttt{Example} class representation:
        \begin{itemize}
          \item Header: tag length is 2 bytes and data length depends on the \texttt{Example} class length.
          \item Data:
                \begin{itemize}
                  \item tag value is \texttt{[0x01, 0x03]} where:
                        \begin{itemize}
                          \item  \texttt{0x01} for \ac{CAP} structure block
                          \item \texttt{0x03} points package \texttt{0x03} (here
                                \texttt{sample} package)
                        \end{itemize}
                        
                  \item data contains the binary representation of \texttt{Example} class.
                \end{itemize}
        \end{itemize}

  \item \texttt{static\_field} block value:
        \begin{itemize}
          \item Header: tag length is 3 bytes and data length is 5 bytes.
          \item Data:
                \begin{itemize}
                  \item tag value is \texttt{[0x02, 0x03, 0x00]} where:
                        \begin{itemize}
                          \item  \texttt{0x02} for static field block
                          \item \texttt{0x03} points package \texttt{0x03} (here
                                \texttt{sample} package)
                          \item \texttt{0x00} static field 0
                        \end{itemize}

                  \item data contains \texttt{[FIELD\_TYPE\_ARRAY\_BYTE, 0x01, 0x02, 0x03, 0x04]}
                \end{itemize}
        \end{itemize}

  \item \texttt{class\_field1} block value:
        \begin{itemize}
          \item Header: tag length is 4 bytes and data length is 3 bytes.
          \item Data:
                \begin{itemize}
                  \item tag value is \texttt{[0x03, 0x03, 0x00, 0x00]} where:
                        \begin{itemize}
                          \item  \texttt{0x03} for applet field block
                          \item \texttt{0x03} points package \texttt{0x03} (here
                                \texttt{sample} package)
                          \item \texttt{0x00} class 0 (here \texttt{Example} class) of package \texttt{sample}
                          \item \texttt{0x00} class field 0
                        \end{itemize}

                  \item data contains \texttt{[FIELD\_TYPE\_SHORT, 0x45, 0x67]}
                \end{itemize}
        \end{itemize}

  \item \texttt{class\_field2} block value:
  \begin{itemize}
    \item Header: tag length is 4 bytes. Data length depends on the
          \texttt{Object} instance size. In our \ac{API} implementation,
          \texttt{Object} class has no field. Therefore, there instance size is
          0. So, the data length is 1 byte.
    \item Data: tag value is \texttt{[0x03, 0x03, 0x00, 0x01]} where:
    \begin{itemize}
      \item \texttt{0x03} for applet field block
      \item \texttt{0x03} points package \texttt{0x03} (here
      \texttt{sample} package)
      \item \texttt{0x00} class 0 (here \texttt{Example} class) of package \texttt{sample}
      \item \texttt{0x01} class field 1
      \item data contains \texttt{[FIELD\_TYPE\_OBJECT]}
    \end{itemize}
  \end{itemize}

\end{enumerate}

% The presented \ac{JCVM} memory organization was evaluated on an ARM board closer
% to evaluate the relevance of our solution. We obtain results which confirm that our
% proposal can be generalized to improve the security of the \ac{JCVM} platform.

% Eventually, we have enough information to generate the initial state. In
% \autoref{sec:experimental-results}, we will generate the initial state for a specific
% target where the Flash memory organization requires to adapt the data location.

\section{Experimental Results}
\label{sec:experimental-results}

For our experimentation, we target an STM32F401RE microcontroller. This
microcontroller embeds an ARM Cortex M4 with \SI{512}{\kilo\byte} Flash
memory. The STM32F401RE Flash memory is split into 8 sectors of 16, 16, 16, 16,
64, 128, 128 and \SI{128}{\kilo\byte} in one bank.\newline

Our Java-source code is based on the Java Card \ac{API} implementations,
GlobalPlatform manager and SmartPGP applet, an open-source GPG
implementation\footnote{This implementation is the one available there:
  \url{https://github.com/ANSSI-FR/SmartPGP}.}. The Java-source code is composed
of 146 Java files split into 23 packages with \SI{8040}{} Java lines of code.
Building the entire Java-source code from \autoref{fig:rommask_generation}
part~\tikz[baseline=0]{\node[circle, dashed, draw, scale=0.75, anchor=base,
  fill=yellow!20] {\textbf{1}};} gives around \SI{51}{\kilo\byte} of data with:

\begin{itemize}
  \item A packages table where each installed packages are listed: \SI{8}{Bytes};

  \item Applications (\ac{API}, GloablPlatform manager and applet) code: \SI{51198}{Bytes};

  \item Non-null static fields initial values: \SI{0}{Byte}. Our
    implementation has not non-null static fields initial values. Static object
    fields need a construct method, called by the \ac{JCRE}, to be initialized.

\end{itemize}

The compilation process does neither optimize the code nor compress it. This
will require future work to obtain a more compact binary output.\newline

Regarding the data size to write, two possibilities are open to us. On the one
hand, we can split data through each sector regarding their size and the data
size to write. This approach may create defragmentation over the sector. On the
other hand, we can write all data in the same sector and rely on the target
\ac{OS} to optimize data location during the \ac{JCVM} lifetime. We choose the
latter solution.

So, at least a \SI{64}{\kilo\byte} sector has to store the entire binary blob
where all \ac{API}, GlobalPlatform manager and applet are to write. We
arbitrarily selected the sector 5. Such as the file system was designed, the
location of data inside the Flash memory is not important. Indeed, the target
\ac{OS} must parse the memory to find all valid blocks during the hash table
generation. \newline

After generating the Java Card initialization image, as introduced in
\autoref{sec:flash-block} and in \autoref{fig:rommask_generation}
part~\tikz[baseline=0]{\node[circle, dashed, draw, scale=0.75, anchor=base,
  fill=orange!20] {\textbf{3}};}, we will need to set the sector 5 of the
STM32F401RE Flash memory. So this can happen, our initial state generator
outputs an Intel Hex file format. This file is commonly used for programming
microcontroller and memory\footnote{The complete file format explanation is
  available on Wikipedia \url{https://en.wikipedia.org/wiki/Intel_HEX}}.\newline

To setup the STM32F401RE Flash memory, we use
\texttt{openocd}\footnote{\texttt{openocd} is available here
  \url{http://openocd.org/}} with the command:
\begin{center}
\texttt{flash write\_image erase flash.hex}
\end{center}

where \texttt{flash.hex} is the file given by the initial state generator.\newline

With the initial state, the \texttt{jni.h} file is also generated
(\SI{403}{lines} length). This file contains the \texttt{callJCNativeMethod}
function. An overview of this file is listed in  \autoref{fig:jni}.

\begin{lstlisting}[language=C,
  numbers=left,
caption={An overview of the \texttt{jni.h} file.}, label={fig:jni},]
/* STARTING METHOD PARAMETERS */
#define STARTING_JAVACARD_PACKAGE 0x01 (*\label{lst:jni:c1}*)
#define STARTING_JAVACARD_CLASS   0x00 (*\label{lst:jni:c2}*)
#define STARTING_JAVACARD_METHOD  0x01 (*\label{lst:jni:c3}*)

/**
 * Java types to provide by the JCVM implementation:
 *   - jref_t: for Java Card reference
 *   - jshort_t: for Java Card short value
 *   - jbyte_t: for Java Card byte value
 *   - jbool_t: for Java Card boolean value
 *   - jint_t: for Java Card integer value
 */

/* METHOD SIGNATURES TO IMPLEMENT */
extern jshort_t
NativeImplementation_arrayCopyRepack(jref_t,jshort_t,jshort_t,jref_t,jshort_t);(*\label{lst:jni:extern}*)
// ...

#define NATIVEIMPLEMENTATION_ARRAYCOPYREPACK         0x0000
// ...
#define NATIVEIMPLEMENTATION_MAKETRANSIENTINTARRAY   0x0021

void callJCNativeMethod(Context& context, jshort_t index) { (*\label{lst:jni:call}*)
  Stack& stack = context.getStack();
  Heap& heap = context.getHeap();
  switch(index) {
    case NATIVEIMPLEMENTATION_ARRAYCOPYREPACK: { (*\label{lst:jni:call_m1_begin}*)
      jshort_t param_04 = stack.pop_Short();
      jref_t   param_03 = stack.pop_Reference();
      jshort_t param_02 = stack.pop_Short();
      jshort_t param_01 = stack.pop_Short();
      jref_t   param_00 = stack.pop_Reference();

      stack.push_Short
         (NativeImplementation_arrayCopyRepack
              (param_00, param_01, param_02,
                         param_03, param_04);
      break; /* ... */ }}} (*\label{lst:jni:call_m1_end}*)

\end{lstlisting}

\autoref{fig:jni}, lines~\ref{lst:jni:c1} to~\ref{lst:jni:c3}, contains a tuple
(package, class, method) which defines the entry point of the first Java Card
method called by the \ac{JCRE} on startup. This method initialized the Java Card
platform and call the GlobalPlatform daemon.\newline

The native method dispatcher, line~\ref{lst:jni:call}, is a switch-case
statement where the native method
\texttt{arrayCopyRepack()} is called
(line~\ref{lst:jni:call_m1_begin} to~\ref{lst:jni:call_m1_end}). As introduced
in the line~\ref{lst:jni:extern}, the
\texttt{arrayCopyRepack()} method is declared as a
C-\texttt{extern} method and must be implemented by the \ac{JCVM} developers. In
our \ac{API} implementation, 34 native methods must be implemented.

\section{Conclusion and Future Works}\label{sec:concl-future-works}

In this article, we introduced a design proposal to organize \ac{NVM} memory
where every data required for running \ac{JCVM} implementation and applets are
stored. The \ac{JCVM} memory organization is not studying by the state of the art. To
improve the security of \ac{JCVM} implementation, every part of the Java Card
platform should be studied to have an open design where the scientific community
can contribute.

This work is partial but it is a new step for designing a \ac{JCVM}
implementation. The next step consists in studying how the whole \ac{JCVM} based
on this \ac{JCVM} memory organization and the \ac{OS} introduced by Bouffard
\etal{}~\cite{conf/sstic/bouffard18} can be secured designed.\newline

Another possible future enhancement would be to optimize the generated data.
Indeed, the current Java Card initialization image is based on the result of
Java compiler. This compiler does not optimize the built bytecodes. Code
optimization may reduce the built code size and, \textit{in fine}, reduce the
size of the resulting binary output. This could allow ones to remove, for
instance, unreachable or useless fragments of code. Security of this
optimization must also be studied.

Also based on the code optimization, code compression can be another approach to
have a smaller data. Code compression replaces several frequently used pack of
instructions by another one. Future research work could study this compression
to obtain a generic and optimized compressing method.

% Acronym list
\begin{acronym}
  \acro{APDU} {Application Protocol Data Unit}
  \acro{API}  {Application Programming Interface}
  \acro{RAM}  {Random Access Memory}
  \acro{ID}   {Identifier}
  \acro{JNI}  {Java Native Interface}
  \acro{BCV}  {Bytecode Verifier}
  \acro{CAP}  {Converted APplet}
  \acro{JCA}  {Java Card Assembly}
  \acro{CFG}  {Control Flow Graph}
  \acro{JCRE} {Java Card Runtime Environment}
  \acro{JCVM} {Java Card Virtual Machine}
  \acro{JVM}  {Java Virtual Machine}
  \acro{JNI}  {Java Native Interface}
  \acro{VM}   {Virtual Machine}
  \acro{MMU}  {Memory Management Unit}
  \acro{IOMMU}{Input-Output Memory Management Unit}
  \acro{SUT}  {System Under Test}
  \acro{HCP}  {Host Controller Protocol}
  \acro{SWP}  {Single Wire Protocol}
  \acro{NFC}  {Near Field Communication}
  \acro{NVM}  {Non-Volatile Memory}
  \acro{MPU}  {Memory Protection Unit}
  \acro{XN}   {eXecute Never}
  \acro{AP}   {Access Permissions}
  \acro{IP}   {Intellectual Property}
  \acro{OS}   {Operating System}
  \acro{CPU}  {Central Processing Unit}
  \acro{syscall}{system call}
  \acro{DoS} {Denial of Service}
  \acro{CFI} {Control Flow Integrity}
  \acro{COTS}{Commercially available Off-The-Shelf}
  \acro{BNF} {Backus\--Naur form}
  \acro{SIO} {Shareable Interface Object}
  \acro{AID} {Application Identifier}
  \acro{HTTP} {Hypertext Transfer Protocol}
  \acro{CVM} {Cardholder Verification Method}
\end{acronym}

\bibliographystyle{splncs04}
\bibliography{article}

\begin{thebibliography}{10}
\providecommand{\url}[1]{\texttt{#1}}
\providecommand{\urlprefix}{URL }
\providecommand{\doi}[1]{https://doi.org/#1}

\bibitem{conf/cardis/BarbuDH11}
Barbu, G., Duc, G., Hoogvorst, P.: {Java Card Operand Stack: Fault Attacks,
  Combined Attacks and Countermeasures}. In: Prouff  \cite{conf/cardis/2011},
  pp. 297--313. \doi{10.1007/978-3-642-27257-8\_19}

\bibitem{conf/secrypt/BarbuHD12}
Barbu, G., Hoogvorst, P., Duc, G.: {Tampering with Java Card Exceptions -- The
  Exception Proves the Rule}. In: Samarati, P., Lou, W., Zhou, J. (eds.)
  {SECRYPT} 2012 -- Proceedings of the International Conference on Security and
  Cryptography, Rome, Italy, 24-27 July, 2012, {SECRYPT} is part of {ICETE} --
  The International Joint Conference on e-Business and Telecommunications. pp.
  55--63. SciTePress (2012)

\bibitem{conf/cardis/BarbuTG10}
Barbu, G., Thiebeauld, H., Guerin, V.: Attacks on java card 3.0 combining fault
  and logical attacks. In: Gollmann, D., Lanet, J.L., Iguchi-Cartigny, J.
  (eds.) Smart Card Research and Advanced Application, 9th {IFIP} {WG} 8.8/11.2
  International Conference, {CARDIS} 2010, Passau, Germany, April 14-16, 2010.
  Proceedings. Lecture Notes in Computer Science, vol.~6035, pp. 148--163.
  Springer (2010). \doi{10.1007/978-3-642-12510-2\_11}

\bibitem{conf/acsac/BenadjilaMRTT19}
Benadjila, R., Michelizza, A., Renard, M., Thierry, P., Trebuchet, P.: {WooKey:
  designing a trusted and efficient {USB} device}. In: Balenson, D. (ed.)
  Proceedings of the 35th Annual Computer Security Applications Conference,
  {ACSAC} 2019, San Juan, PR, USA, December 09-13, 2019. pp. 673--686. {ACM}
  (2019). \doi{10.1145/3359789.3359802}

\bibitem{conf/javacard/BieberCMGLWZ00}
Bieber, P., Cazin, J., Marouani, A.E., Girard, P., Lanet, J.L., Wiels, V.,
  Zanon, G.: {The {PACAP} Prototype: {A} Tool for Detecting Java Card Illegal
  Flow}. In: Attali, I., Jensen, T.P. (eds.) {Java on Smart Cards: Programming
  and Security, First International Workshop, JavaCard 2000, Cannes, France,
  September 14, 2000, Revised Papers}. Lecture Notes in Computer Science,
  vol.~2041, pp. 25--37. Springer (2000). \doi{10.1007/3-540-45165-X\_3}

\bibitem{conf/sstic/bouffard18}
Bouffard, G., Gaspard, L.: {Hardening a Java Card Virtual Machine
  Implementation with the MPU}. Symposium sur la sécurité des technologies de
  l'information et des communications (SSTIC)  (2018)

\bibitem{conf/cardis/BouffardIL11}
Bouffard, G., Iguchi-Cartigny, J., Lanet, J.L.: Combined software and hardware
  attacks on the java card control flow. In: Prouff  \cite{conf/cardis/2011},
  pp. 283--296. \doi{10.1007/978-3-642-27257-8\_18}

\bibitem{conf/cardis/BouffardLLL14}
Bouffard, G., Lackner, M., Lanet, J.L., Loinig, J.: {Heap ... Hop! Heap Is Also
  Vulnerable}. Lecture Notes in Computer Science, vol.~8968, pp. 18--31.
  Springer (2014). \doi{10.1007/978-3-319-16763-3\_2}

\bibitem{journals/virology/BouffardL14}
Bouffard, G., Lanet, J.L.: {Reversing the operating system of a Java based
  smart card}. Journal Computer Virology and Hacking Techniques
  \textbf{10}(4),  239--253 (2014). \doi{10.1007/s11416-014-0218-7}

\bibitem{journals/compsec/BouffardL15}
Bouffard, G., Lanet, J.L.: {The ultimate control flow transfer in a Java based
  smart card}. Computers {\&} Security  \textbf{50},  33--46 (2015).
  \doi{10.1016/j.cose.2015.01.004}

\bibitem{conf/sensys/DaiNH04}
Dai, H., Neufeld, M., Han, R.: {ELF:} an efficient log-structured flash file
  system for micro sensor nodes. In: Stankovic, J.A., Arora, A., Govindan, R.
  (eds.) Proceedings of the 2nd International Conference on Embedded Networked
  Sensor Systems, SenSys 2004, Baltimore, MD, USA, November 3-5, 2004. pp.
  176--187. {ACM} (2004). \doi{10.1145/1031495.1031516}

\bibitem{conf/sstic/Dubreuil16}
Dubreuil, J.: Java card security, software and combined attacks. Symposium sur
  la sécurité des technologies de l'information et des communications (SSTIC)
   (2016)

\bibitem{journals/ijsse/DubreuilBTL13}
Dubreuil, J., Bouffard, G., Thampi, B.N., Lanet, J.L.: {Mitigating Type
  Confusion on Java Card}. {IJSSE}  \textbf{4}(2),  19--39 (2013).
  \doi{10.4018/jsse.2013040102}

\bibitem{phd/dufray03}
Dufay, G.: {V\'erification formelle de la plate-forme Java Card}. Ph.D. thesis,
  INRIA Sophia-Antipolis, France (Dec 2003)

\bibitem{book/spec/etsi_ts_102_226}
ETSI: {Smart Cards; Remote APDU structure for UICC based applications (Release
  12)}. ETSI, etsi ts 102 226 edn. (Feb 2015)

\bibitem{conf/aasri/farissiALM13}
Farissi, I., Azizi, M., Lanet, J.L., Moussaoui, M.: {Neural Network Vs.
  Bayesian Network to Detect Java Card Mutants}. AASRI Procedia  \textbf{4} (12
  2013). \doi{10.1016/j.aasri.2013.10.021}

\bibitem{conf/cardis/Faugeron13}
Faugeron, E.: {Manipulating the Frame Information with an Underflow Attack}.
  In: Francillon, A., Rohatgi, P. (eds.) {Smart Card Research and Advanced
  Applications - 12th International Conference, {CARDIS} 2013, Berlin, Germany,
  November 27-29, 2013. Revised Selected Papers}. Lecture Notes in Computer
  Science, vol.~8419, pp. 140--151. Springer (2013).
  \doi{10.1007/978-3-319-08302-5\_10}

\bibitem{book/spec/gp_spe11}
GlobalPlatform: {Card Specification}. GlobalPlatform Inc., 2.2.1 edn. (Jan
  2011)

\bibitem{journals/spe/GuptaN16}
Gupta, K., Nandivada, V.K.: {Lexical state analyzer for JavaCC grammars}.
  Journal of Software - Practice and Experience (SPE)  \textbf{46}(6),
  751--765 (2016). \doi{10.1002/spe.2322}

\bibitem{conf/sarssi/HamadoucheBLDNMR12}
Hamadouche, S., Bouffard, G., Lanet, J.L., Dorsemaine, B., Nouhant, B.,
  Magloire, A., Reygnaud, A.: {Subverting Byte Code Linker service to
  characterize Java Card API}. In: {Seventh Conference on Network and
  Information Systems Security (SAR-SSI)}. pp. 75--81 (May~22rd to 25th 2012)

\bibitem{journals/istr/HamadoucheL13}
Hamadouche, S., Lanet, J.L.: {Virus in a smart card: Myth or reality?} Journal
  of Information Security and Applications  \textbf{18}(2-3),  130--137 (2013).
  \doi{10.1016/j.jisa.2013.08.005}

\bibitem{journals/jce/IdrissiBLH17}
Idrissi, N.E.J.E., Bouffard, G., Lanet, J.L., Hajji, S.E.: {Trust can be
  misplaced}. Journal of Cryptographic Engineering  \textbf{7}(1),  21--34
  (2017). \doi{10.1007/s13389-016-0142-5}

\bibitem{journals/virology/Iguchi-CartignyL10}
Iguchi-Cartigny, J., Lanet, J.L.: {Developing a Trojan applets in a smart
  card}. Journal in Computer Virology  \textbf{6}(4),  343--351 (2010).
  \doi{10.1007/s11416-009-0135-3}

\bibitem{conf/cardis/LacknerBLWS12}
Lackner, M., Berlach, R., Loinig, J., Weiss, R., Steger, C.: {Towards the
  Hardware Accelerated Defensive Virtual Machine - Type and Bound Protection}.
  In: Mangard  \cite{conf/cardis/2012}, pp. 1--15.
  \doi{10.1007/978-3-642-37288-9\_1}

\bibitem{conf/hipeac/LacknerBWS14}
Lackner, M., Berlach, R., Weiss, R., Steger, C.: {Countering type confusion and
  buffer overflow attacks on Java smart cards by data type sensitive
  obfuscation}. In: Knoop, J., Salapura, V., Koren, I., Pelosi, G. (eds.)
  {Proceedings of the First Workshop on Cryptography and Security in Computing
  Systems, CS2@HiPEAC 2014, Vienna, Austria, January 20, 2014}. pp. 19--24.
  {ACM} (2014). \doi{10.1145/2556315.2556317}

\bibitem{masterthesis/Lafer14}
Lafer, M.: Design and Implementation of a Java Card Operating System for Design
  Space Exploration on Different Platforms. Master's thesis, Graz University of
  Technology (Jan 2014)

\bibitem{conf/cardis/Lancia12}
Lancia, J.: {Java Card Combined Attacks with Localization-Agnostic Fault
  Injection}. In: Mangard  \cite{conf/cardis/2012}, pp. 31--45.
  \doi{10.1007/978-3-642-37288-9\_3}

\bibitem{conf/cardis/LanciaB15}
Lancia, J., Bouffard, G.: {Java Card Virtual Machine Compromising from a
  Bytecode Verified Applet}. In: Homma, N., Medwed, M. (eds.) {Smart Card
  Research and Advanced Applications - 14th International Conference, {CARDIS}
  2015, Bochum, Germany, November 4-6, 2015. Revised Selected Papers}. Lecture
  Notes in Computer Science, vol.~9514, pp. 75--88. Springer (2015).
  \doi{10.1007/978-3-319-31271-2\_5}

\bibitem{conf/cardis/2012}
Mangard, S. (ed.): {Smart Card Research and Advanced Applications - 11th
  International Conference, {CARDIS} 2012, Graz, Austria, November 28-30, 2012,
  Revised Selected Papers}, Lecture Notes in Computer Science, vol.~7771.
  Springer (2013). \doi{10.1007/978-3-642-37288-9}

\bibitem{journals/ijisec/MesbahLM19}
Mesbah, A., Lanet, J.L., Mezghiche, M.: Reverse engineering java card and
  vulnerability exploitation: a shortcut to {ROM}. Int. J. Inf. Sec.
  \textbf{18}(1),  85--100 (2019). \doi{10.1007/s10207-018-0401-9}

\bibitem{conf/cardis/MostowskiP08}
Mostowski, W., Poll, E.: {Malicious Code on Java Card Smartcards: Attacks and
  Countermeasures}. In: Grimaud, G., Standaert, F.X. (eds.) {Smart Card
  Research and Advanced Applications, 8th {IFIP} {WG} 8.8/11.2 International
  Conference, {CARDIS} 2008, London, UK, September 8-11, 2008. Proceedings}.
  Lecture Notes in Computer Science, vol.~5189, pp. 1--16. Springer (2008).
  \doi{10.1007/978-3-540-85893-5\_1}

\bibitem{press_release_nxp_2020}
NXP: {NXP Launches New Java Card-based Operating System to Expand
  Multi-Application Services in the Secure Identification Market}. NXP Press
  Release website  (Sep 2020),
  \url{https://media.nxp.com/news-releases/news-release-details/nxp-launches-new-java-card-based-operating-system-expand-multi}

\bibitem{book/oracle15/dev_kit}
Oracle: {Java Card 3 Platform, Development Kit 3.0.5}. Oracle (Sep 2011)

\bibitem{book/oracle15/jcre_spec}
Oracle: {Java Card 3 Platform, Runtime Environment Specification, Classic
  Edition 3.0.5}. Oracle (Sep 2011)

\bibitem{book/oracle15/api_spec}
Oracle: {Java Card 3 Platform, API, Classic Edition 3.0.5}. Oracle (2015)

\bibitem{book/oracle15/jcvm_spec}
Oracle: {Java Card 3 Platform, Virtual Machine Specification, Classic Edition
  3.0.5}. Oracle (2015)

\bibitem{ppjavacard}
Oracle: Java card\texttrademark{} protection profile collection (2017)

\bibitem{prevost2006application}
Prevost, S., Sachdeva, K.: {Application code integrity check during virtual
  machine runtime} (Mar~2 2006), uS Patent App. 10/929,221

\bibitem{conf/cardis/2011}
Prouff, E. (ed.): {Smart Card Research and Advanced Applications - 10th {IFIP}
  {WG} 8.8/11.2 International Conference, {CARDIS} 2011, Leuven, Belgium,
  September 14-16, 2011, Revised Selected Papers}, Lecture Notes in Computer
  Science, vol.~7079. Springer (2011). \doi{10.1007/978-3-642-27257-8}

\bibitem{conf/dbsec/RazafindralamboBL12}
Razafindralambo, T., Bouffard, G., Lanet, J.L.: {A Friendly Framework for
  Hidding fault enabled virus for Java Based Smartcard}. In: Cuppens-Boulahia,
  N., Cuppens, F., Garc{\'{\i}}a-Alfaro, J. (eds.) {DBSec 2012, Paris, France,
  July 11-13,2012. Proceedings}. Lecture Notes in Computer Science, vol.~7371,
  pp. 122--128. Springer (2012). \doi{10.1007/978-3-642-31540-4\_10}

\bibitem{conf/snds/RazafindralamboBTL12}
Razafindralambo, T., Bouffard, G., Thampi, B.N., Lanet, J.L.: {A Dynamic Syntax
  Interpretation for Java Based Smart Card to Mitigate Logical Attacks}. In:
  Thampi, S.M., Zomaya, A.Y., Strufe, T., Calero, J.M.A., Thomas, T. (eds.)
  {Recent Trends in Computer Networks and Distributed Systems Security -
  International Conference, {SNDS} 2012, Trivandrum, India, October 11-12,
  2012. Proceedings}. Communications in Computer and Information Science,
  vol.~335, pp. 185--194. Springer (2012). \doi{10.1007/978-3-642-34135-9\_19}

\bibitem{conf/fm/SchierlSHR09}
Schierl, A., Schellhorn, G., Haneberg, D., Reif, W.: Abstract specification of
  the {UBIFS} file system for flash memory. In: Cavalcanti, A., Dams, D. (eds.)
  {FM} 2009: Formal Methods, Second World Congress, Eindhoven, The Netherlands,
  November 2-6, 2009. Proceedings. Lecture Notes in Computer Science,
  vol.~5850, pp. 190--206. Springer (2009). \doi{10.1007/978-3-642-05089-3\_13}

\bibitem{conf/fgit/SereIL10}
S{\'{e}}r{\'{e}}, A.A.K., Iguchi-Cartigny, J., Lanet, J.L.: {Checking the Paths
  to Identify Mutant Application on Embedded Systems}. In: Kim, T.H., Lee,
  Y.H., Kang, B.H., Slezak, D. (eds.) {Future Generation Information Technology
  - Second International Conference, {FGIT} 2010, Jeju Island, Korea, December
  13-15, 2010. Proceedings}. Lecture Notes in Computer Science, vol.~6485, pp.
  459--468. Springer (2010). \doi{10.1007/978-3-642-17569-5\_45}

\bibitem{sun2014research}
Sun, F., Zhang, F.x.: Research an improvement of {YAFFS} file system. Comput.
  Eng  \textbf{34},  258--261 (2014)

\bibitem{woodhouse2001jffs}
Woodhouse, D.: Jffs: The journalling flash file system. In: Ottawa linux
  symposium. vol.~2001 (2001)

\bibitem{masterthesis/Zelle15}
Zelle, M.: Design and Implementation of a Hardware Supported Memory Protection
  for the Java Card Firewall. Master's thesis, Graz University of Technology
  (Apr 2015)

\end{thebibliography}

\end{document}